\documentclass[12pt,american]{article}

\usepackage[T1]{fontenc} 
\usepackage{babel} 
\usepackage{lmodern} 

\usepackage{geometry}
\usepackage{setspace} 
\usepackage[authoryear]{natbib} 

\usepackage{amsmath,amssymb,amsthm,thmtools,mathtools,mathrsfs} 
\usepackage{enumitem,algorithm2e} 

\usepackage{tabularx, booktabs} 

\usepackage[unicode=true,pdfusetitle,
 bookmarks=true,bookmarksnumbered=false,bookmarksopen=false,
 breaklinks=true,pdfborder={0 0 0},pdfborderstyle={},backref=false,colorlinks=true]
 {hyperref}
\hypersetup{linkcolor=blue, citecolor=blue, urlcolor=blue}

\usepackage[normalem]{ulem} 
\usepackage{xcolor} 
\usepackage{pifont} 
\usepackage{bbding} 
\usepackage{comment} 

\newcommand{\noun}[1]{\textsc{#1}}

\newlength{\lyxlabelwidth}      

\theoremstyle{remark}
\newtheorem{claim}{Claim}
\theoremstyle{plain}
\newtheorem{prop}{Proposition}
\newtheorem{lem}{Lemma}
\newtheorem{thm}{Theorem}
\newtheorem{cor}{Corollary}
\theoremstyle{definition}
\newtheorem{example}{Example}
\newtheorem{property}{Property}

\let\variant\relax

\declaretheorem[
  name={Property~\variant{\textsuperscript{*}}}, 
  style=definition,
  numbered=no,
]{property*}

\newenvironment{rproperty}[1]
  {\newcommand\variant{\ref*{#1}}\begin{property*}}
  {\end{property*}}
  
\declaretheorem[
  name={Property~\variant{\textsuperscript{\#}}},
  style=definition,
  numbered=no,
]{property-q}

\declaretheorem[
  name={Property~\variant{\textsuperscript{\textdagger}}},
  style=definition,
  numbered=no,
]{property-v}

\newenvironment{elabeling}[2][]
	{\settowidth{\lyxlabelwidth}{#2}
		\begin{description}[font=\normalfont,style=sameline,
			leftmargin=\lyxlabelwidth,#1]}
	{\end{description}}

\newlist{casenv}{enumerate}{4}
\setlist[casenv]{leftmargin=*,align=left,widest={iiii}}
\setlist[casenv,1]{label={{\itshape\ Case} \arabic*.},ref=\arabic*}
\setlist[casenv,2]{label={{\itshape\ Case} \roman*.},ref=\roman*}
\setlist[casenv,3]{label={{\itshape\ Case\ \alph*.}},ref=\alph*}
\setlist[casenv,4]{label={{\itshape\ Case} \arabic*.},ref=\arabic*}

\newenvironment{subproof}[1][\proofname]{
  
  \begin{proof}[#1]
}{
  \end{proof}
}

\begin{document}
\title{Axiomatic Characterizations of Draft Rules
}
\author{Jacob Coreno\textsuperscript{a,}\thanks{Corresponding author.} \and Ivan Balbuzanov\textsuperscript{b}}
\maketitle

\renewcommand{\thefootnote}{\alph{footnote}}

\footnotetext[1]{Faculty of Business and Economics, University of Lausanne, Internef 530, 1015 Lausanne, Switzerland. \texttt{jacob.coreno@unil.ch}}
\footnotetext[2]{Department of Economics, Level 4, Building 105, University of Melbourne VIC 3010, Australia. \texttt{ivan.balbuzanov@unimelb.edu.au}}

\begin{abstract}
Drafts are sequential round-robin allocation procedures
for distributing heterogeneous and indivisible objects among agents
subject to some priority order (e.g., allocating players' contract
rights to teams in professional sports leagues). Agents report ordinal
preferences over objects and bundles are partially ordered by pairwise
comparison. We provide a simple characterization of \emph{draft
rules}: they are the only allocation rules that are \emph{respectful
of a priority (RP)}, \emph{envy-free up to one object (EF1)},
\emph{non-wasteful (NW)}, and \emph{resource monotonic (RM)}.
RP and EF1 are crucial for competitive balance in sports leagues.
We also prove two related impossibility theorems showing that the competitive-balance axioms RP and EF1 are generally incompatible with \emph{strategy-proofness}. Nevertheless, draft
rules satisfy \emph{maxmin strategy-proofness}. If agents may declare some objects
unacceptable, then draft rules are characterized by RP, EF1, NW, and
RM, in conjunction with \emph{individual rationality} and \emph{truncation invariance}.
\end{abstract}



\section{Introduction}
\renewcommand{\thefootnote}{\arabic{footnote}}
A draft is a natural procedure for distributing heterogeneous and indivisible resources: agents take turns to select one object at a time, in an order specified by some priority, until all objects are gone (or all agents are satiated). Drafts are used in a variety of multiple-object assignment problems, such as the allocation of university courses \citep{budish2012multi} and political positions \citep{bram04,olea05}, divorce settlement \citep{will99}, and estate division \citep{heath_2018}.\footnote{Drafts typically see use in settings that value fair object allocation. See \url{https://www.onlinedraft.com/} for a variety of other applications.} Undoubtedly, the most prominent application---and the most economically important one---is the annual draft of professional sports leagues, through which existing teams are assigned the rights to sign new players. 

Drafts are used by the ``Big Four'' North American sports leagues, the National Football League (NFL), Major League Baseball (MLB), the National Basketball Association (NBA), and the National Hockey League, as well as other sports leagues in North America, Australia, and elsewhere. These leagues generate substantial revenues, attract significant interest and fan engagement from audiences, both live and TV,\footnote{While NFL's Super Bowl is typically the most-watched TV program in the US, the NFL draft, held annually during the off-season, is a huge event in its own right. In April 2022, the three-day event amassed an average TV audience of 5.2 million, with over 10 million viewers tuning in for the first round. The NBA draft attracts similarly high viewership.} and the teams that compete in them are extremely valuable.\footnote{Each of the North American Big Four is among the five leagues with the highest revenue in the world. The four of them enjoy a combined annual revenue of over \$54 billion USD. The NFL's 32 franchises have a combined market value of around \$228 billion USD, making it the most valuable league in the world \citep{bade25}.}

A unique attribute of professional sports leagues is that ``competitors
must be of approximately equal `size' if any are to be successful''
\citep[p.\ 242]{rottenberg1956baseball}. The conventional wisdom is that closely contested matches
with uncertain outcomes generate more spectator interest and, ultimately,
larger profits. In line with this conventional wisdom, studies, starting with \citet{schm01} and \citet{humphreys2002alternative}, have uncovered a positive empirical relationship between measures of \emph{competitive balance} (i.e., minimizing the variance in league members' competitive abilities) and average attendance.

Within this context, league officials and franchise owners see the draft as an important pillar of competitive balance (e.g., \citealp{grier1994rookie}). Specifically, by giving the worst performing teams higher priority and the first choices in the draft, the best new talent is reserved for weaker teams, making them more competitive in subsequent
seasons.\footnote{Another reason that the draft may appeal to team officials is that it minimizes inter-team competition for the most talented rookies and prevents bidding wars over the most desirable incoming players. A player that is new to a league may sign a contract only with the team that drafted him or her.} Indeed, studies, such as \cite{grier1994rookie} and \cite{butler1995competitive}, empirically demonstrate the
essential role that the draft plays in maintaining and promoting competitive balance. Yet there has been limited progress toward a solid theoretical understanding of how the properties of the draft, when viewed as a formal matching mechanism, promote competitive balance.

This paper analyzes drafts and competitive balance using an axiomatic approach. We focus on \emph{draft rules}, which are centralized allocation
rules that closely resemble the decentralized drafts used by professional sports leagues.\footnote{In a \emph{(centralized) allocation rule}, agents simultaneously report
their preferences directly to a central authority, and the allocation
is determined based on the reported preferences according to some
rule or procedure. In contrast, a \emph{(decentralized) draft} is
an extensive-form mechanism in which agents choose objects sequentially,
and the allocation is determined by the sequence of selections made
by the agents. Our results pertain to centralized draft rules, which
are simpler to analyze than their decentralized counterparts. Nevertheless,
our results can also be applied to decentralized drafts, provided
that teams always select their most-preferred player whenever they
are called to pick.}\textsuperscript{,}\footnote{Note that this is a pure object-allocation problem and not a two-sided matching market. The problem is to allocate players' contract rights to teams rather than to match teams and players together as in centralized labor markets, such as the NRMP. Players' preferences over the teams are not used to determine the final allocation.} More precisely, a draft rule allocates recruits to the teams over
several rounds; each team is assigned one recruit per round, in an
order specified by some priority ordering. We assume that each team has a total order over recruits, which generates a partial order over sets of recruits via pairwise comparison.\footnote{\label{fn:Complementarities}That is, a team prefers set $A$ over set $B$ if there exists a one-to-one function from $B$ to $A$ that maps each element of $B$ into a more-preferred element of $A$. Partial orders from pairwise comparison are similar to responsive preferences. The difference is that we do not require the existence of a complete preference order over the sets of recruits. Our assumption rules out complicated
preferences over allocations, such as those exhibiting complementarities
(e.g., ``we want quarterback A only if we also get wide receiver
B'').}
Our main contribution is, to the best of our knowledge, the first axiomatic characterization of draft rules: they are the only allocation rules satisfying \emph{respect for priority (RP)}, \emph{envy-freeness up to one object (EF1)}, \emph{resource monotonicity (RM)}, and \emph{non-wastefulness (NW)}.

Our characterization gives an axiomatic justification for the use of draft rules in professional sports leagues. The first two properties, RP and EF1, are closely tied to  competitive
balance. RP requires that there exists a priority ordering
over the teams such that each team prefers its own set of recruits
to the set of recruits assigned to \emph{any other team with lower
priority}. RP can be viewed as a weak fairness property (as in \citealp{svensson1994queue}) that guarantees the absence of justified envy \citep{abdulkadirouglu2003school}. The class of allocation rules satisfying RP is large and
includes draft rules (where, as is typically the case in sports leagues,
the relevant priority is determined by reverse finishing order in
the previous season---or by a weighted lottery based on this order)
as well as \emph{serial dictatorships}.\footnote{A serial dictatorship works as follows: the first team is assigned
its most-preferred set of recruits; the second team is assigned its
most-preferred set of recruits among the pool of remaining recruits,
and so on.} Serial dictatorships are, however, highly inequitable. In particular, they can
provide too much support for the weaker teams, making them far too
strong in the following seasons.\footnote{\label{fn:EF1 failure}For example, the 2003 NBA draft is known for
producing one of the most talented cohorts of all time. Four of the
top five picks, namely LeBron James, Carmelo Anthony, Chris Bosh, and
Dwyane Wade, were NBA All-Stars. LeBron James, Dwyane Wade, and Chris
Bosh became all-time scoring leaders for their respective teams after
just their first seven seasons. The trio, who became known as the
``Big Three'', united at the Miami Heat between 2010 and 2014; they led the Heat to the NBA Finals
in each of their four seasons together, winning back-to-back championships
in 2012 and 2013. If the NBA had instead used a serial dictatorship,
the Cleveland Cavaliers---the worst performing team in the 2002-03
season---could have selected multiple members of the ``Big Three'' on
their single turn, helping the team to immediate success.}

EF1 requires that every team prefers its own set of recruits to the
recruits assigned to \emph{any other team} after at
most one player is removed from the other team's set. To see why drafts satisfy EF1, consider a draft in which team $i$ selects players $a_1, \dots, a_k$ (in that order), while team $j$ selects $b_1, \dots, b_k$. If each of team $i$'s picks precede $j$'s in the same round, then $j$ may prefer $i$'s assignment to its own---i.e., $j$ may \emph{envy} $i$. However, if $a_1$ is removed from $i$'s assigned bundle, then $j$ will cease envying $i$: it prefers its own assignment $\{b_1,b_2, \dots, b_k\}$ over $\{a_2, \dots, a_k\}$, because at each round $\ell$ the choice $b_\ell$ was revealed preferred to $a_{\ell+1}$.

EF1 rules out serial dictatorships,
together with their associated problems: by limiting the extent to
which low-priority teams can envy their high-priority rivals, EF1
ensures that high-priority teams are not favored too heavily. Intuitively,
EF1 allocation rules can prevent drastic swings in team quality across
seasons, allowing for some stability in year-to-year team rankings;
they can also limit the incentives to ``tank,'' relative to other
allocation rules satisfying RP.\footnote{\emph{Tanking} is the practice of sports teams making suboptimal roster or tactical decision with the goal of losing matches and obtaining higher priority in the next season's draft. If the NBA had instead used a serial dictatorship in 2003, the prospect of drafting multiple members of the ``Big Three'' could have tempted many teams to tank.}

A non-wasteful allocation rule assigns all available players.\footnote{\label{efficiency footnote}In the working-paper version of this article \citep{core24}, we show that the draft satisfies a stronger efficiency criterion: it always selects an allocation that is Pareto efficient with respect to the teams' (partial) preferences over sets of recruits. As the allocation of players to teams permits externalities (making one team ``better off'' by giving it better players makes defeating it in play more difficult, thus indirectly making all other teams worse off), we de-emphasize Pareto efficiency here. Efficiency remains desirable in other settings such as university course allocation, as well as property- and estate-division problems, where these externalities are not typically present. Note also that draft rules are not efficient with respect to \emph{each} complete extension of the agents' partial preferences over bundles \citep{hatfield2009strategy}. This is why some authors call this property \emph{Pareto possibility} (e.g., \citealt{bram03}) or \emph{possible Pareto efficiency} (e.g., \citealp{aziz2019efficient}) instead.} RM captures a notion of solidarity among the teams: it requires that, whenever the set of available recruits grows, all teams become weakly better off.

Although draft rules satisfy many appealing properties, they possess
a fundamental defect: they are not \emph{strategy-proof}
\emph{(SP)} as a team may benefit from picking insincerely if
it knows that its most-preferred recruit will not be picked by other
teams. It is therefore natural to ask whether there
exists a strategy-proof allocation rule satisfying the competitive
balance properties, RP and EF1. This question is answered in the negative by two
impossibility results. The main insight from Theorems~\ref{thm:impossibility 1} and~\ref{thm:impossibility 2} is that
every allocation rule satisfying the competitive balance properties, RP and EF1,
possesses one of two defects: it is either manipulable or inefficient.
Hence, a sports league can remedy the draft's deficiencies only if
it is willing to introduce another deficiency---one which is arguably
more severe.

While draft rules are not strategy-proof, there are compelling reasons
for teams to report their preferences truthfully. First, no team can
profitably misrepresent its preferences at any problem instance in
which all teams share the same preferences. More generally,
Theorem~\ref{thm:Maxminimizer} shows that, if teams have \emph{additive}
preferences,\footnote{An agent has additive preferences if its preferences over bundles
are represented by an additive utility function. See footnote~\ref{fn:Additive utility function}
for a definition of additive utility functions.} then truthful reporting is a utility maxminimizer in the revelation
game associated with any draft rule. In other words, draft rules are
\emph{maxmin strategy-proof (MSP)}. Truth-telling is optimal
when agents are maxmin expected utility maximizers and they are uncertain (in the Knightian sense)
about the other teams' preferences \citep{gilboa1989maxmin}. As a corollary of Theorem~\ref{thm:Maxminimizer}, it follows that, even though there may exist a strategic manipulation that benefits a given team, any such manipulations are non-obvious and, therefore, draft rules are not obviously manipulable in the sense of \citet{troy20}.

The rest of this paper is organized as follows. Section~\ref{Brief overview of related literature} gives an overview of the related literature. Section~\ref{Model} introduces the standard model, in which allocation rules are formally defined, and we discuss how to extend agents' preferences over objects to incomplete preferences over bundles (e.g., \emph{sets of recruits}) using the pairwise dominance extension (see, for example, \citealp{brams1979prisoners}). Section~\ref{Properties} defines several properties of allocation rules and highlights some useful relationships between the properties. Section~\ref{Draft Rules} provides a formal definition of draft rules. Section~\ref{Results} states the main results of the paper. Section~\ref{s:unacc_ext} considers a more general preference domain, in which agents may deem certain objects unacceptable, and presents a version of the main characterization result. Section~\ref{sec:conclusion} concludes. All proofs are postponed to Appendix \ref{sec:appendix}.

\subsection{Related Literature\label{Brief overview of related literature}}

\citet{rottenberg1956baseball} and \citet{neale1964peculiar}
were among the first to highlight the importance of competitive balance in sports leagues. \citet{borland2003demand} provide an excellent review of the literature, and they discuss the connection between competitive balance and spectator demand. 
\citet{humphreys2002alternative} introduces a long-run
measure of competitive balance and shows that this measure is a significant
determinant of total annual attendance in MLB. Several important papers
are devoted to empirically evaluating the impact that drafts have
on competitive balance. For instance, \citet{grier1994rookie} show
that earlier picks in the NFL draft are associated with better performance
in subsequent seasons, and \citet{butler1995competitive} shows that
the MLB draft has significantly reduced the within-season dispersion
of winning percentages.

\citet{kohler1971class} provide one of the earliest theoretical studies of drafts.
They give an algorithm for constructing
a subgame perfect Nash equilibrium in a decentralized two-team drafting game. \citet{brams1979prisoners} show that, with only two teams, the equilibrium allocation is Pareto efficient
with respect to pairwise comparison; however, this is not necessarily
true when there are three or more teams (see also \citealp{manea2007serial}). For a (centralized) draft rule with arbitrarily many teams, \citet{aziz2017equilibria} show that the sincere outcome---the allocation that results under truthful reporting---can be supported by a Nash equilibrium. \citet{brams2005efficient}
show that an allocation is Pareto efficient with respect to pairwise comparison if and only if it is the sincere outcome of a draft rule associated with some picking sequence. When agents have additive preferences, \citet{caragiannis2019unreasonable} point out that draft rules satisfy a version of EF1; furthermore, each pure strategy Nash equilibrium induces an allocation that is EF1 with respect to the agents' true preferences (\citealp{amanatidis2024allocating}; see also \citealp{amanatidis2024round}). Given the well-established fact that draft rules are manipulable
 (e.g., \citealp{kohler1971class}; \citealp{hatfield2009strategy}), \citet{bouveret2011general, bouveret2014manipulating}
study the problem of computing optimal manipulations, while \citet{aziz2017complexity}
consider the computational complexity of such problems.

\citet{budish2012multi} study a version of the draft used to allocate courses to students at the Harvard Business School. The mechanism, which is sometimes known as the \emph{snake draft}, is based on a randomly drawn initial priority that reverses after each round. In a related paper, \citet{aziz2015possible} characterize the \emph{allocations} that are outcomes of sequential-allocation rules associated with natural picking sequences, such as ``strict alternation'' (drafts), ``balanced alternation'' (snake drafts), and other ``recursively balanced'' sequences. \citet{biro2022serial} study \emph{Single-Serial} rules, which are equivalent to draft rules with agent-specific quotas. Single-Serial rules are shown to be \emph{ig-Pareto-efficient}, equivalent to Pareto efficiency in our model, as well as \emph{truncation-proof} (see also \citealp{biro2022balanced}). \citet{caspari2022booster} analyzes a related allocation rule, called a \emph{booster draft}. In a booster draft with $n$ agents, the objects are partitioned into $m$ sets of size $n$ (called \emph{boosters}), and each agent picks once from each of the $m$ sets. Since an agent's choice from one booster does not impact the set of objects available to her in another booster, the rule is strategy-proof. However, booster drafts are not efficient.\footnote{In a related model with endowments, \citet{papai2003strategyproof} introduces the \emph{segmented trading cycles} rules in which agents trade objects, one-for-one, within restricted market segments. Similar to booster drafts, these rules are strategy-proof but inefficient.}

Our characterization results complement existing characterizations of alternative allocation rules for the multiple-object assignment problem. For example, \citet{papai2001strategyproof}, \citet{ehlers2003coalitional}, and
\citet{hatfield2009strategy} show, on various preference domains,
that \emph{sequential dictatorships}\footnote{In a sequential dictatorship, a first dictator is assigned
her most-preferred bundle from the entire set of objects. Then, a second
dictator, whose identity may depend on the first dictator and her assigned
bundle, chooses his most-preferred bundle from the set of remaining
objects; the third dictator is determined by the previous dictators and their
assigned bundles, and so on. If the order of dictators is the same
for all preference profiles, then the sequential dictatorship is a serial dictatorship.} are characterized by appropriate notions
of efficiency, strategy-proofness, and non-bossiness. In
each of those papers, imposing further requirements (e.g., ``neutrality''
in \citealp{hatfield2009strategy}) pins down the class of serial
dictatorships (see also \citealp{papai2000strategyproof} and \citealp{klaus2001strategy}).

The properties we consider are mostly familiar, but they have been
adapted or extended to our setting. Many mechanisms allocate objects
to agents on the basis of priorities. For the class of \emph{single-unit} problems, where each agent is assigned a single object, \cite{svensson1994queue} considers a version of RP (which he calls \emph{weak fairness}) and constructs a mechanism satisfying RP, SP, and Pareto efficiency on the full preference domain. \citet[Lemma 1]{ergin2000consistency} shows that this property, together with Pareto efficiency, fully characterizes serial dictatorships in the single-unit setting with strict preferences. Our version of RP is a natural extension of \citeauthor{svensson1994queue}'s (\citeyear{svensson1994queue}) property to multiple-object problems. Another related property is \emph{no justified envy (NJE)}, which is a desideratum for school-choice mechanisms \citep{abdulkadirouglu2003school}: both properties forbid envy towards an agent with lower priority. The difference is that NJE arises in problems with unit demand and local (i.e., school-specific), rather than global, priority orders.

EF1 has its roots in the work of \citet{lipton2004approximately},
but the version we use is adapted from \citet{budish2011combinatorial}.
\citet{budish2011combinatorial} considers complete preferences over
bundles, whereas our EF1 is based on the pairwise dominance extension.
Our efficiency criterion, Pareto efficiency with respect to pairwise
dominance, is identical to that of \citet{brams1979prisoners} and
\citet{brams2005efficient}; it is weaker than the efficiency
criteria considered by \citet{ehlers2003coalitional}
and \citet{hatfield2009strategy}, where agents are assumed to have
complete responsive preferences over bundles. \citet{ehlers2003coalitional} consider a version of resource monotonicity; their version is more permissive than our RM, since they do not require
that each finite subset of potential objects is an admissible set
of available objects. Truncation invariance is common in settings
where objects to be allocated may be unacceptable to some agents. It is used by \citet{ehlers2014strategy} to characterize Deferred Acceptance, and by \citet{hashimoto2014two} to characterize
the Probabilistic Serial rule.

Our impossibility theorems parallel existing results that highlight
the tension between three ideal desiderata: efficiency, fairness, and
incentive compatibility. For example, our Theorem~\ref{thm:impossibility 2}
is similar to Proposition~3 of \citet{budish2011combinatorial}, which
states that strategy-proofness is incompatible with efficiency and envy-freeness
up to one object in his model. \citet{caspari2022booster} proves a related result, which says that, with a fixed set of available objects, strategy-proofness is incompatible with efficiency and \emph{envy-freeness up to $k$ objects}, for some $k$ that depends on the number of objects available.

\section{Model and definitions}

\subsection{Model\label{Model}}

We consider the problem of allocating bundles of heterogeneous and
indivisible objects to members of a fixed population of $n$ agents. Let $N\coloneqq\left\{ 1,\dots,n\right\} $
denote the fixed set of \emph{agents}, where $n\geq2$. The set of objects
available to the agents can change. Let $\mathbb{O}$ denote the set
of potential \emph{objects}, where $\mathbb{O}$ is assumed to be
countably infinite. Let ${\cal X}\coloneqq\left\{ X\subseteq\mathbb{O}\mid 0<\lvert X\rvert<\infty\right\} $
denote the family of sets of \emph{available objects}, which are finite nonempty
subsets of $\mathbb{O}$.

A \emph{(strict) preference} $\succeq_{i}$ for agent $i$ is a linear order on $\mathbb{O}$---that is, a complete,
transitive and antisymmetric binary relation on $\mathbb{O}$. We
write $\succ_{i}$ for the asymmetric part of $\succeq_{i}$, i.e.,
for all $x,y\in\mathbb{O}$, $x\succ_{i}y$ if and only if $x\succeq_{i}y$
and $x\neq y$. Let ${\cal R}$ denote the set of all preferences.
The set of all preference profiles $\succeq=\left(\succeq_{i}\right)_{i\in N}$
is denoted by ${\cal R}^{N}$. A \emph{bundle} is a finite (possibly empty) subset of $\mathbb{O}$. The set of all bundles is denoted ${\cal B}\coloneqq{\cal X}\cup\left\{ \emptyset\right\} $.

Given a set $X\in\mathcal{B}$ and a preference relation
$\succeq_{i}\in{\cal R}$, we denote by $\succeq_{i}\mid_{X}$ the
restriction of $\succeq_{i}$ to the set $X$. If, for example, $X=\left\{ x_{1},x_{2},\dots,x_{k}\right\} $
and $x_{1}\succ_{i}x_{2}\succ_{i}\cdots\succ_{i}x_{k}$, then we will
sometimes use shorthand notation such as $X=\left\{ x_{1}\succ_{i}x_{2}\succ_{i}\cdots\succ_{i}x_{k}\right\} $
or $\succeq_{i}\mid_{X}=x_{1},x_{2},\dots,x_{k}$. If, in addition, for each $x \in X$ and $y\in\mathbb{O}\setminus X$ we have
$x\succ_{i}y$, then we may write $\succeq_{i}=x_{1},x_{2},\dots,x_{k},\dots$. If $X\neq\emptyset$, then $\operatorname{top}_{\succeq_{i}}\left(X\right)$
denotes the top-ranked object according to $\succeq_{i}$; that is,
$\operatorname{top}_{\succeq_{i}}\left(X\right)=x$ if and only if $x\in X$ and
$x\succeq_{i}y$ for all $y\in X$. Moreover, given $k \in \{0,1,\dots,\lvert X\rvert\}$,
let $\operatorname{top}_{\succeq_{i}}\left(X,k\right)$ denote the set consisting of the $k$ top-ranked
objects in $X$ (which is empty if $k=0$).

An \emph{allocation} $A=\left(A_{i}\right)_{i\in N}$ is a profile
of disjoint bundles, where $A_{i}$ is the bundle of objects assigned
to agent $i$. Let ${\cal A}$ denote the set of all allocations.
For each $X\in{\cal X}$, ${\cal A}\left(X\right)$ denotes the subset
of ${\cal A}$ consisting of all allocations $A=\left(A_{i}\right)_{i\in N}$
satisfying $\bigcup_{i\in N}A_{i}\subseteq X$. A \emph{problem} is
a pair $\left(\succeq,X\right)\in{\cal R}^{N}\times{\cal X}$. An
\emph{allocation rule} is a function $\varphi:{\cal R}^{N}\times{\cal X}\to{\cal A}$
such that $\varphi\left(\succeq,X\right)\in{\cal A}\left(X\right)$.\footnote{To simplify notation, we allow agents to express preferences over
all objects in $\mathbb{O}$, even though the set $X$ of available
objects is a proper subset of $\mathbb{O}$.} We emphasize that the allocation rules considered in this paper are
defined over preference profiles on the set of \emph{objects}, rather
than an alternative domain consisting of preference profiles on the
set of \emph{bundles}. The latter domain is much richer but would impose significant complexity costs to participants.

To facilitate meaningful comparisons of allocation rules, agents'
preferences are extended from objects to bundles via pairwise dominance \citep{brams1979prisoners}. A bundle $T$ is pairwise dominated by bundle $S$ for $i$ if for every element of $T$ there corresponds a distinct element in $S$ that is weakly preferred by $i$. Formally, the \emph{pairwise dominance extension $\succeq_{i}^{PD}$}
of a preference relation $\succeq_{i}$ is defined as follows: for
all $S,T\in{\cal B}$, $S\succeq_{i}^{PD}T$ if and only if there
is an injective function $\mu:T\to S$ such that $\mu\left(x\right)\succeq_{i}x$
for each $x\in T$. The relation $\succeq_{i}^{PD}$
is a partial order on ${\cal B}$. Denote by $\succ_{i}^{PD}$ the asymmetric part of $\succeq_{i}^{PD}$. Throughout the paper, we interpret $\succeq_{i}^{PD}$ as agent $i$'s (incomplete) ranking over
${\cal B}$ when her preference relation is $\succeq_{i}$.

The pairwise dominance extension $\succeq^{PD}_i$ is related to standard assumptions commonly imposed on an agent's complete preferences over bundles. It is \emph{monotone} (i.e., for each $S,T\in{\cal B}$, $S \supseteq T$ implies $S \succeq_{i}^{PD} T$), reflecting the assumption that agents' preferences exhibit non-satiation or that ``more is better.''\footnote{In the context of sports drafts, contract rights typically have option value and free disposal: a team can choose to not sign players it has drafted and instead retain their contract rights as a tradeable asset.} Note that $S\succeq_i^{PD}\emptyset$ is vacuously true for every $S\in\mathcal{B}$. Furthermore, $\succeq_i^{PD}$ equals the intersection of all additive preferences consistent with $\succeq_i$, where a preference over bundles is \emph{additive} if it can be represented by an additive utility function\footnote{\label{fn:Additive utility function}A utility function $u_{i}:{\cal B}\to\mathbb{R}_{+}$ is \emph{additive} if, for all $X\in{\cal B}$, $u_{i}\left(X\right)=\sum_{x\in X}u_{i}\left(\left\{ x\right\} \right)$, where it is understood that $u_{i}\left(\emptyset\right)=0$ and $u_i(\{x\}) > 0$ for all $x \in \mathbb{O}$.} and \emph{consistent with $\succeq_i$} if it agrees with $\succeq_i$ in comparisons of singleton bundles. Additionally, $\succeq^{PD}_i$ also equals the intersection of all monotone \emph{responsive} preferences consistent with $\succeq_i$ \citep{roth85}. 
As discussed in footnote~\ref{fn:Complementarities}, the assumption
that agents compare bundles by pairwise dominance rules out preferences
exhibiting ``complementarities.''\footnote{For instance, if $x,y,z$ are
	distinct objects and $P_{i}$ is a strict total order on ${\cal B}$,
	then it is conceivable that $\left\{ x\right\} P_{i}\left\{ y\right\} P_{i}\left\{ z\right\} $
	and $\left\{ x,z\right\} P_{i}\left\{ x,y\right\} $. This can happen if, for example, $x$ and $y$ are (mutually substitutable) NFL quarterbacks, while
	$z$ is a wide receiver, a position that is complementary to the quarterback. However, if $x\succ_{i}y\succ_{i}z$, then agent $i$ necessarily
	prefers $\left\{ x,y\right\} $ to $\left\{ x,z\right\}$ in our
	model.}

\subsection{Properties of Allocation Rules\label{Properties}}

\subsubsection{Fairness and Competitive Balance}\label{sec:fairness}

Given an allocation $A$ and a preference profile $\succeq$, $A_{i}\succeq_{i}^{PD}A_{j}$ for some agents $i$ and $j$ guarantees that, for any additive utility function $u_i$ consistent with $\succeq_i$, $i$ derives higher utility from $A_i$ than from $A_j$ or, intuitively, $i$ cannot envy $j$. Otherwise, if $A_{i}\nsucceq_{i}^{PD}A_{j}$, there exists $u_i$ such that $u_i(A_j)>u_i(A_i)$ or, in other words, $i$ may possibly envy $j$. For simplicity, we say that $i$ envies $j$ if $A_{i}\nsucceq_{i}^{PD}A_{j}$.

The requirement that no agent envies any other agent (\emph{envy-freeness}) would be far too demanding in our setting. It is easy to see that any envy-free allocation rule is wasteful: if all agents have the same preferences, it allocates the empty bundle to each agent.

Furthermore, there are many practical settings in which envy is \emph{desirable}. For instance, sports leagues may seek to allocate the best new players to weaker teams in order to redress competitive imbalances. An envy-free allocation rule would do little to correct these issues---insofar as each team's preference over players is based on the players' ability to contribute to the team's victory probability, an envy-free rule is likely to perpetuate the dominance of already successful teams. Similarly, students with higher exam scores enjoy higher priority for school seats \citep{abdulkadirouglu2003school}.

While such practices violate envy-freeness, they are generally considered fair. To model these situations, we introduce a priority ordering over the agents as follows. A \emph{priority} is a strict linear order $\pi$ on $N$. We say that agent $i$ has higher priority than agent $j$ if $i\mathrel{\pi} j$. The following property requires that no agent envies an agent with lower priority.

\begin{property}
\label{prop:RP}
An allocation rule $\varphi$ is \emph{respectful
of the priority $\pi$} \emph{(RP-$\pi$)} if, for any problem $\left(\succeq,X\right)$
and any agent $i\in N$,
\[
\varphi_{i}\left(\succeq,X\right)\succeq_{i}^{PD}\varphi_{j}\left(\succeq,X\right)\text{ for each }j\in N\text{ such that }i\mathrel{\pi} j.
\]
An allocation rule $\varphi$ is \emph{respectful of a priority (RP)}
if there exists a priority $\pi$ such that $\varphi$ satisfies RP-$\pi$.
\end{property}

If $\varphi$ is an allocation rule that respects
the priority $\pi$, then it is possible that an agent envies
another agent with higher priority. The second weakening of envy-freeness
requires that the degree of any envy is not ``too large'' in the
following sense: if agent $i$ envies agent $j$, then this envy can
be eliminated by removing exactly one object from agent $j$'s bundle.
It is easy to see that EF1 is logically independent from RP.

\begin{property}
\label{prop:EF1}
An allocation rule $\varphi$ is \emph{envy-free
up to one object (EF1)} if, for any problem $\left(\succeq,X\right)$
and agents $i,j\in N$, there exists $S\subseteq\varphi_{j}\left(\succeq,X\right)$
such that $\lvert S\rvert\leq1$ and $\varphi_{i}\left(\succeq,X\right)\succeq_{i}^{PD}\varphi_{j}\left(\succeq,X\right)\setminus S$.
\end{property}

\subsubsection{Non-wastefulness}

The next property, non-wastefulness (NW), is a minimal requirement for efficiency. It simply says that all available objects are assigned.

\begin{property}
\label{prop:NW}
An allocation rule $\varphi$ is \emph{non-wasteful
(NW)} if, for any problem $\left(\succeq,X\right)$, $\bigcup_{i\in N}\varphi_i\left(\succeq,X\right)=X$.
\end{property}

If an allocation rule violates NW, then we call it \emph{wasteful}. Non-wastefulness is mild but desirable in most object-allocation settings. For example, in divorce settlements or estate divisions, assets are often universally desirable, so leaving an item unassigned would be inefficient. Non-wastefulness remains relatively innocuous even in the context of sports drafts, where a team can hold a player's contract rights without necessarily signing them---that is, free disposal applies. Section~\ref{s:unacc_ext} considers an extension that drops this assumption and allows agents to find some objects unacceptable.\footnote{The online appendix considers a related extension, in which agents have possibly heterogeneous quotas on their bundle size. These may arise in settings where the maximum squad size binds or in drafts that have a fixed number of rounds. After the allocation-rule properties are appropriately redefined to account for the binding quotas, our main characterization result, Theorem~\ref{thm:main characterization}, carries through unchanged. The online appendix also shows that Theorem~\ref{thm:main characterization} can likewise be extended to allow for objects with multiple copies if each agent receives no more than one copy of each object. This occurs in university course allocation, for instance.}

\subsubsection{Resource Monotonicity}

Recall that an allocation rule $\varphi$ specifies an allocation
$\varphi\left(\succeq,X\right)$ for every problem $\left(\succeq,X\right)$,
where $X$ is any set of available objects drawn from ${\cal X}$.
The following property requires, loosely speaking, that monotonic
changes in the set of available objects are associated with monotonic
changes in the welfare of every agent. Specifically, whenever the
set of available objects grows (shrinks), all agents receive a weakly
better (worse) bundle. \citet{chun1988monotonicity} and \citet{moulin1988can} offered some of the earliest analyses of rules satisfying this property.

\begin{property}
\label{prop:RM}
An allocation rule $\varphi$ is \emph{resource monotonic
(RM)} if, for any $X,X'\in{\cal X}$ and $\succeq\in{\cal R}^{N}$,
\[
X\supseteq X'\implies\varphi_{i}\left(\succeq,X\right)\succeq_{i}^{PD}\varphi_{i}\left(\succeq,X'\right)\text{ for all }i\in N.
\]
\end{property}

While RM is of secondary importance to sports leagues, it is certainly appealing in other applications such as university course allocation, as well as property- and estate-division problems. For the former one, RM guarantees that a university does not reduce any student's happiness with their course selection by adding more courses or increasing their capacity. For the latter two, RM guarantees that the parties do not have an incentive to destroy assets.

\subsection{Draft Rules\label{Draft Rules}}

A \emph{picking sequence} is a function $f:\mathbb{N}\to N$, where $\mathbb{N}=\left\{ 1,2,3,\dots\right\}$. The
\emph{sequential-allocation rule associated with $f$} is the allocation rule $\varphi^{f}:{\cal R}^{N}\times{\cal X}\to{\cal A}$
that assigns agents their favorite remaining object, one at a time,
in the order prescribed by $f$. More precisely, $\varphi^{f}$ associates with each problem $\left(\succeq,X\right)$ the allocation $\varphi^{f}\left(\succeq,X\right)\in{\cal A}\left(X\right)$
determined by the following procedure. 

\paragraph*{\rule{1\columnwidth}{1pt}}

\paragraph{Algorithm 1:\label{par:Algorithm 1}}

\noun{Draft$\left(\succeq,X,f\right)$.}
\begin{elabeling}{00.00.0000}
\item [{\emph{Input:}}] A problem $\left(\succeq,X\right)$ and a picking
sequence $f$.
\item [{\emph{Output:}}] An allocation $\varphi^{f}\left(\succeq,X\right)\in{\cal A}\left(X\right)$.
\end{elabeling}
\begin{enumerate}
\item Define the sequence $\left(s_{k}\right)_{k=1}^{\lvert X\rvert}$ of
\emph{selections} recursively. Set:
\begin{enumerate}
\item $s_{1}=\operatorname{top}_{\succeq_{f\left(1\right)}}\left(X\right)$.
\item for $k=2,\dots,\lvert X\rvert$, $s_{k}=\operatorname{top}_{\succeq_{f\left(k\right)}}\left(X\setminus\left\{s_{1},\dots,s_{k-1}\right\} \right)$.
\end{enumerate}
\item For each $i\in N$, set $\varphi_{i}^{f}\left(\succeq,X\right)=\left\{ s_{k}\mid f\left(k\right)=i,1\leq k\leq \lvert X\rvert\right\} $.
\item Return $\varphi^{f}\left(\succeq,X\right)=\left(\varphi_{i}^{f}\left(\succeq,X\right)\right)_{i\in N}$.
\end{enumerate}
\rule{1\columnwidth}{1pt}

Given a priority $\pi$, let $f^{\pi}:\mathbb{N}\to N$ denote the
\emph{picking sequence associated with $\pi$}, defined by
\[
\left(f^{\pi}\left(k\right)\right)_{k\in\mathbb{N}}=\left(\underset{\text{round 1}}{\underbrace{\pi_1,\pi_2,\dots,\pi_n}},\underset{\text{round 2}}{\underbrace{\pi_1,\pi_2,\dots,\pi_n}},\dots\right),
\]
where $\pi_1\coloneqq\operatorname{top}_{\pi} N$ is the agent with the highest priority according to $\pi$ and generally $\pi_i\coloneqq\operatorname{top}_{\pi} N\setminus\{\pi_1,\ldots,\pi_{i-1}\}$ is the agent with the $i$th highest priority according to $\pi$.
The \emph{draft rule associated with $\pi$} is the allocation rule
$\varphi^{\pi}\coloneqq\varphi^{f^{\pi}}$. If an allocation rule
$\varphi$ satisfies $\varphi=\varphi^{\pi}$ for some priority $\pi$,
then we call $\varphi$ a \emph{draft rule}.

\section{Main results\label{Results}}

\subsection{Characterization Result}\label{sss:main_char_result}

The following result is the main characterization result of the paper. It says that $\varphi^{\pi}$ is the \emph{unique} allocation rule satisfying RP-$\pi$, EF1, NW, and RM.

\begin{thm}\label{thm:main characterization}
An allocation rule satisfies RP-$\pi$, EF1, NW, and RM if and only if it equals $\varphi^\pi$.
\end{thm}

The proof of the theorem establishes a stronger uniqueness result, using the following weakenings of RP and EF1, which concern only the relative sizes of the agents' assigned bundles. They suffice for the majority of our results. Weak RP requires that higher-priority agents have larger bundles, while weak EF1 requires that the bundle sizes of any two agents differ by at most one.

\begin{property}
\label{prop:WRP}
An allocation rule $\varphi$ is \emph{weakly respectful
of the priority $\pi$ (wRP-$\pi$)} if, for any problem $\left(\succeq,X\right)$
and any agent $i\in N$,
\[
\lvert\varphi_{i}\left(\succeq,X\right)\rvert\geq\lvert\varphi_{j}\left(\succeq,X\right)\rvert\text{ for each }j\in N\text{ such that }i\mathrel{\pi}j.
\]
An allocation rule $\varphi$ is \emph{weakly respectful of a priority (wRP)}
if there exists a priority $\pi$ such that $\varphi$ satisfies wRP-$\pi$.
\end{property}

\begin{property}
\label{prop:WEF1}
An allocation rule $\varphi$ is \emph{weakly envy-free up to one object (wEF1)} if, for any problem $\left(\succeq,X\right)$ and any agents $i,j\in N$,
\[
\lvert \varphi_{j}\left(\succeq,X\right)\rvert \leq\lvert \varphi_{i}\left(\succeq,X\right)\rvert +1.
\]
\end{property}

The proof of Theorem~\ref{thm:main characterization} shows that (i) $\varphi^\pi$ satisfies RP-$\pi$, EF1, NW, and RM, and (ii) if $\varphi$ satisfies wRP-$\pi$, wEF1, NW, and RM, then it equals $\varphi^\pi$. The following example illustrates the proof's main ideas.

\begin{example}

Suppose there are three agents, $N=\left\{ 1,2,3\right\} $, and four
available objects, say $X=\left\{ a,b,c,d\right\} $. Consider a
preference profile $\succeq\in{\cal R}^{N}$ satisfying
\begin{gather*}
		\succeq_{1}\mid_{X} = a,b,c,d; \quad \succeq_{2}\mid_{X} = c,d,b,a; \quad \succeq_{3}\mid_{X} = a,d,c,b.
\end{gather*}
Letting $\pi$ denote the identity priority (i.e., $\pi_i=i$
for each $i\in N$), we will show that any allocation rule $\varphi$
satisfying NW, wRP-$\pi$, wEF1, and RM must satisfy $\varphi\left(\succeq,X\right)=\varphi^{\pi}\left(\succeq,X\right)=\left(\left\{ a,b\right\} ,\left\{c\right\} ,\left\{d\right\} \right)$.
The main idea of the proof is to inductively consider a sequence of sets of available objects: starting from the empty set, we add
objects from $X$, one at a time, according to the order in which they
are selected in the draft procedure \noun{Draft}$\left(\succeq,X,f^{\pi}\right)$.
We then show that, for each of the subsets of
available objects $S_{k}$, where $\emptyset\subsetneq S_{1}\subsetneq S_{2}\subsetneq\dots\subsetneq S_{\lvert X\rvert}=X$,
one has $\varphi\left(\succeq,S_{k}\right)=\varphi^{\pi}\left(\succeq,S_{k}\right)$.

The sequence of selections associated with the draft procedure \noun{Draft}$\left(\succeq,X,f^{\pi}\right)$
is $\left(s_{k}\right)_{k=1}^{\lvert X\rvert}=\left(a,c,d,b\right)$.
For each $k=1,\dots,\lvert X\rvert$, we define $S_{k}=\left\{ s_{1},\dots,s_{k}\right\} $:
\begin{gather*}
S_{1}=\left\{a\right\} ,\quad S_{2}=\left\{a,c\right\} ,\quad S_{3}=\left\{a,c,d\right\} ,\quad S_{4}=\left\{a,b,c,d\right\}.
\end{gather*}
We make the following observations:
\begin{enumerate}
\item $\varphi\left(\succeq,S_{1}\right)=\left(\left\{a\right\} ,\emptyset,\emptyset\right)$.

By NW, object $a$ must be assigned to some agent; by wRP-$\pi$, it
must be assigned to agent~$1$.

\item $\varphi\left(\succeq,S_{2}\right)=\left(\left\{a\right\} ,\left\{c\right\} ,\emptyset\right)$.

By NW, both objects must be assigned. If object~$a$ is assigned to anyone other than agent~1, RM would be violated as 1 would not be weakly better off with respect to $\succeq^{PD}_1$ after the addition of $c$. Therefore, $a$ must be assigned to 1. See also Lemma~\ref{lem:RM lemma} in Appendix \ref{sec:appendix}, which codifies a general version of this observation that is useful for proving Theorem~\ref{thm:main characterization}.

If object~$c$ is assigned to agent~1, wEF1 would be violated with 1 having two more objects than both 2 and 3. If $c$ is assigned to 3, wRP-$\pi$ would be violated as 3's bundle would be larger than 2's bundle. Therefore, $c$ must be assigned to 2. This observation can be generalized to yield Lemma~\ref{lem:bundle size lemma} in Appendix \ref{sec:appendix}, which is another important stepping stone towards the proof of Theorem~\ref{thm:main characterization}.

\item $\varphi\left(\succeq,S_{3}\right)=\left(\left\{a\right\} ,\left\{c\right\} ,\left\{d\right\} \right)$.

By NW, all objects must be assigned. Similarly to the above, if object~$a$ is assigned to anyone other than agent~1 or object~$c$ is assigned to anyone other than agent~2, RM would be violated. By wRP-$\pi$, object~$d$ cannot be assigned to 2 and, by wEF1, it cannot be assigned to 1. Therefore, it must be assigned to agent~$3$.

\item $\varphi\left(\succeq,S_{4}\right)=\left(\left\{a,b\right\} ,\left\{c\right\} ,\left\{d\right\} \right)$.

By NW, all objects must be assigned. By RM, objects~$a$, $c$, and~$d$ must
be assigned to agents~1, 2, and~3, respectively. By wRP-$\pi$, object $b$ is assigned to agent $1$.

\end{enumerate}

These observations demonstrate that $\varphi\left(\succeq,S_{k}\right)=\varphi^{\pi}\left(\succeq,S_{k}\right)$
for each $k=1,\dots,\lvert X\rvert$ and, in particular, that $\varphi\left(\succeq,X\right)=\varphi^{\pi}\left(\succeq,X\right)$.
The proof of the general result extends the preceding argument to
show that $\varphi\left(\succeq',X'\right)=\varphi^{\pi}\left(\succeq',X'\right)$
for \emph{any} problem $\left(\succeq',X'\right)$.\hfill{}$\diamond$
\end{example}

Theorem~\ref{thm:main characterization} shows that the draft rule $\varphi^{\pi}$ is fully characterized by RP-$\pi$, EF1, RM, and NW. The four properties in Theorem~\ref{thm:main characterization} are independent of one another. To verify that each property is not implied by the remaining three, we give an example of an allocation rule that satisfies the remaining properties but is not a draft rule.

\paragraph{Non-wasteful~(NW).}\label{Null rule}

Let $\varphi$ denote the \emph{null rule}, defined by $\varphi\left(\succeq,X\right)=\left(\emptyset\right)_{i\in N}$
for each problem $\left(\succeq,X\right)$. Then $\varphi$ satisfies
RP-$\pi$ for any priority $\pi$, EF1, and RM, but it violates NW.

\paragraph{Weakly~envy-free~up~to~one~object~(wEF1).}\label{pi-Dictatorship}

Given any priority $\pi$, let $\varphi$ denote the \emph{$\pi$-dictatorship
rule}, defined by $\varphi_{\pi_1}\left(\succeq,X\right)=X$
and $\varphi_{i}\left(\succeq,X\right)=\emptyset$ for each $i\neq\pi_1$.
Then $\varphi$ satisfies NW, RP-$\pi$, and RM,
but not wEF1 (consider any $X$ with $\lvert X\rvert>1$).

\paragraph{Weakly~respectful~of~a~priority~(wRP).}\label{Snake draft}

For any priority $\pi$, let $g^{\pi}$ denote the picking sequence 
\[
\left(g^{\pi}\left(k\right)\right)_{k\in\mathbb{N}}=\left(\underset{\text{round 1}}{\underbrace{\pi_{1},\pi_{2},\dots,\pi_{n}}},\underset{\text{round 2}}{\underbrace{\pi_{n},\pi_{n-1},\dots,\pi_{1}}},\underset{\text{round 3}}{\underbrace{\pi_{1},\pi_{2},\dots,\pi_{n}}},\underset{\text{round 4}}{\underbrace{\pi_{n},\pi_{n-1},\dots,\pi_{1}}},\dots\right).
\]
That is, $g^{\pi}$ orders agents from highest to lowest priority according to $\pi$ in all ``odd rounds,'' and reverses the order in all ``even rounds.'' The associated sequential-allocation rule $\varphi^{g^\pi}$ is the \emph{snake draft} associated with $\pi$. It is easy to see that it satisfies NW, EF1, and RM, but it does not satisfy wRP.

\paragraph{Resource monotonic~(RM).}\label{RM example} Let $\mathbb{O}=\left\{ x_{1},x_{2},\dots\right\} $. Consider
the set $X'=\left\{ x_{1},\dots,x_{n+1}\right\} $ and a preference
profile $\succeq'\in{\cal R}^{N}$ satisfying
\begin{align*}
\succeq'_{1}\mid_{X'} & =x_{1},x_{2},\dots,x_{n+1};\\
\succeq'_{i}\mid_{X'} & =x_{2},\dots,x_{n+1},x_{1},\text{ for all }i\in N\setminus\left\{ 1\right\} .
\end{align*}
Consider a picking sequence $f:\mathbb{N}\to N$ satisfying
$\left(f\left(k\right)\right)_{k=1}^{n+1}=\left(1,1,2,\dots,n\right)$. Then
\[
\varphi_{1}^{f}\left(\succeq',X'\right)=\left\{ x_{1},x_{2}\right\} \quad\text{and}\quad\varphi_{i}^{f}\left(\succeq',X'\right)=\left\{ x_{i+1}\right\} \text{ for all }i\in N\setminus\left\{ 1\right\}.
\] 

Let $\varphi^{\neg{\mathrm{RM}}}$ be the allocation rule given by
\[
\varphi^{\neg{\mathrm{RM}}}\left(\succeq,X\right)=\begin{cases}
\varphi^{f}\left(\succeq',X'\right) & \text{if}~\succeq=\succeq'~\text{and}~X=X';\\
\varphi^{\pi}\left(\succeq,X\right) & \text{otherwise},
\end{cases}
\]
where $\varphi^{\pi}$ is the draft rule associated with the identity priority $\pi$. It is clear that $\varphi^{\neg{\mathrm{RM}}}$ satisfies NW and RP-$\pi$ (and wRP-$\pi$). To show that $\varphi^{\neg{\mathrm{RM}}}$ satisfies EF1, by Theorem~\ref{thm:main characterization} it is enough to consider the allocation at $\varphi^{f}\left(\succeq',X'\right)$. Since $x_{i+1}\succeq'_{i}x_{1}$ for each $i\in N\setminus\left\{ 1\right\} $, we have that EF1 holds. But $\varphi^{\neg{\mathrm{RM}}}$ fails RM, since
\[
\varphi^{\neg{\mathrm{RM}}}_{2}\left(\succeq',X'\setminus\left\{ x_{n+1}\right\} \right)=\varphi_{2}^{\pi}\left(\succeq',X'\setminus\left\{ x_{n+1}\right\} \right)=\left\{ x_{2}\right\} \succ_{2}^{\prime PD}\left\{ x_{3}\right\} =\varphi_{2}^{f}\left(\succeq',X'\right)=\varphi^{\neg{\mathrm{RM}}}_{2}\left(\succeq',X'\right).
\]

\subsection{Impossibility Results and Incentive Properties of the Draft}

We start this section by introducing two incentive compatibility properties.
For a preference profile $\succeq$ and a preference relation $\succeq'_{i}$
of agent $i$, let $\left(\succeq'_{i},\succeq_{-i}\right)$ denote
the preference profile in which agent $i$'s preference is $\succeq'_{i}$
and, for each agent $j\in N\setminus\left\{i\right\} $, agent $j$'s
preference is $\succeq_{j}$. An allocation rule is \emph{weakly strategy-proof} if, by misrepresenting her preferences, no agent can obtain a bundle that is strictly better than the bundle she obtains by reporting her preferences truthfully.

\begin{property}
\label{prop:WSP}
An allocation rule $\varphi$ is \emph{weakly strategy-proof
(wSP)} if, for any problem $\left(\succeq,X\right)$ and $i\in N$,
there is no $\succeq'_{i}\in{\cal R}$ such that $\varphi_{i}\left(\left(\succeq'_{i},\succeq_{-i}\right),X\right)\succ_{i}^{PD}\varphi_{i}\left(\succeq,X\right)$.
\end{property}

This property is a weak version of the usual strategy-proofness property, which would require that truthful reporting guarantees a bundle that is at least as good as any bundle obtainable via misreporting. Its definition is natural for settings in which agents
partially order the set of alternatives (see \citealp{bogomolnaia2001new}
for a classic example where the alternatives are lotteries, partially
ordered by \emph{first-order stochastic dominance}).

Our next property is also weaker than strategy-proofness but it is nevertheless a compelling notion of incentive
compatibility: an allocation rule is \emph{maxmin strategy-proof (MSP)} if truthful reporting is a utility maxminimizer whenever agents have additive
preferences. In other words, truth-telling is optimal when agents
are maxmin expected utility maximizers and face substantial uncertainty about the other agents' preferences \citep{gilboa1989maxmin}. One can show that wSP and MSP are logically independent.

\begin{property}
\label{prop:MSP}
An allocation rule $\varphi$ is \emph{maxmin strategy-proof (MSP)} if, for any $X\in{\cal X}$, $i\in N$,
$\succeq_{i}\in{\cal R}$, 
\begin{equation}
\succeq_{i} \in \underset{\succeq'_{i}\in{\cal R}}{\arg\max}\left[\min_{\succeq'_{-i}\in{\cal R}^{N\setminus \{i\}}}u_{i}\left(\varphi_{i}\left(\left(\succeq'_{i},\succeq'_{-i}\right),X\right)\right)\right]
\end{equation}
for every additive utility function $u_{i}$ consistent with $\succeq_{i}$ (see footnote~\ref{fn:Additive utility function}).
\end{property}

The most concerning deficiency of draft rules is that they are not
strategy-proof. Intuitively, an agent may want to rank an object that is popular with other agents higher in her reported preference order at the expense of an unpopular object that she likes more. The following example illustrates this for the simple case with $n=2$ agents; the example can be extended to larger problems but the general case with $n\geq2$ agents is also implied by Theorem~\ref{thm:impossibility 1} below.

\begin{example}
Suppose there are $n=2$ agents. Without loss of generality, let $\pi$
be the identity priority and consider the allocation rule $\varphi^{\pi}$.
Consider the set $X=\left\{ a,b,c\right\} $ of available objects
and a preference profile $\succeq\in{\cal R}^{N}$ satisfying $\succeq_{1}\mid_{X} =a,b,c$ and $\succeq_{2}\mid_{X}=b,c,a$.
Suppose agent 1 misrepresents her preferences by reporting $\succeq'_{1}\in{\cal R}$,
where $\succeq'_{1}\mid_{X}=b,a,c$. Then
\[
\left\{ a,b\right\} =\varphi_{1}^{\pi}\left(\left(\succeq'_{1},\succeq_{-1}\right),X\right)\succ_{1}^{PD}\varphi_{1}^{\pi}\left(\succeq,X\right)=\left\{ a,c\right\} ,
\]
which violates wSP.\hfill{}$\diamond$
\end{example}

One may ask whether there is a weakly strategy-proof allocation rule
that satisfies the competitive balance properties EF1 and RP,\footnote{Recall that an allocation rule is RP if it is RP-$\pi$ for some priority $\pi$.} making it appealing to sports leagues. The following theorem
shows that any such rule is wasteful. Thus, every
non-wasteful allocation rule satisfying the competitive balance properties
is manipulable.

\begin{thm}
\label{thm:impossibility 1}No allocation rule satisfies
RP, EF1, NW, and wSP.
\end{thm}

One can omit RP in Theorem~\ref{thm:impossibility 1} if NW is replaced
with Pareto efficiency, a more demanding property.

\begin{property}
\label{prop:EFF}
An allocation rule $\varphi$ is \emph{(Pareto) efficient (EFF)} if, for any problem $\left(\succeq,X\right)$, there is no allocation $B\in{\cal A}\left(X\right)$ such that $B\neq \varphi\left(\succeq,X\right)$ and $B_{i}\succeq_{i}^{PD}\varphi_i\left(\succeq,X\right)$ for each $i\in N$.
\end{property}

\begin{thm}
\label{thm:impossibility 2}No allocation rule satisfies
EFF, EF1, and wSP.
\end{thm}

Table~\ref{tab:ruleprops} summarizes the properties of the draft rule and the four rules defined in the axiom-independence examples in Section~\ref{sss:main_char_result}. It demonstrates that EF1, NW, and wSP are indispensable in the statement of Theorem~\ref{thm:impossibility 1}, as are all properties in Theorem~\ref{thm:impossibility 2}. Whether there exists an allocation rule satisfying NW, EF1, and wSP is unclear. In fact, we conjecture that NW, EF1 and wSP are incompatible. Such a result would subsume Theorems~\ref{thm:impossibility 1} and~\ref{thm:impossibility 2}. The allocation rules marked as efficient satisfy EFF because each of them selects, at every problem, the outcome of some sequential-allocation rule for that preference profile, and such outcomes are efficient with respect to pairwise dominance \citep{brams2005efficient}.

\begin{table}[ht]
\centering
	\begin{tabular}{l*{6}{c}}
	\hline
	Allocation Rule & RP-$\pi$ & EF1 & NW & EFF & RM & wSP\\
	\midrule
	$\varphi^\pi$ (Draft rule for $\pi$) & \checkmark & \checkmark & \checkmark & \checkmark & \checkmark & \\
	$\varphi^{g^\pi}$ (Snake draft for $\pi$) & & \checkmark & \checkmark & \checkmark & \checkmark &  \\
	$\pi$-Dictatorship & \checkmark & & \checkmark & \checkmark &  \checkmark & \checkmark \\
	Null rule & \checkmark & \checkmark &  & & \checkmark & \checkmark \\
	$\varphi^{\neg{\mathrm{RM}}}$ & \checkmark & \checkmark & \checkmark & \checkmark &  &  \\
	\bottomrule
	\end{tabular}
\caption{Properties of selected allocation rules}
\label{tab:ruleprops}
\end{table}


The impossibility results above highlight the strong conflict between the
competitive-balance properties and agents' incentives. Nevertheless,
the following result shows that draft rules are maxmin strategy-proof.

\begin{thm}
\label{thm:Maxminimizer}Draft rules are maxmin strategy-proof.
\end{thm}

Theorem~\ref{thm:Maxminimizer} shows that truth-telling is optimal when agents are maxmin expected utility maximizers---this
is the case if, say, agents are uncertainty averse and they face sufficient uncertainty about the other agents' preferences \citep{gilboa1989maxmin}.

\citet{troy20} propose a typology of rules that are not strategy-proof, dividing them into those that are obviously manipulable and those that are not. This categorization is done by comparing the worst- and best-case outcomes of a profitable manipulation to the worst- and best-case outcomes of a truthful report, respectively. More concretely, a draft rule $\varphi^\pi$ is \emph{not obviously manipulable} if for any profitable deviation $\succeq'_i$, we have
\begin{align*}
\min_{\succeq'_{-i}\in{\cal R}^{N\setminus \{i\}}}u_{i}\left(\varphi^\pi_{i}\left(\left(\succeq'_{i},\succeq'_{-i}\right),X\right)\right) &\leq \min_{\succeq'_{-i}\in{\cal R}^{N\setminus \{i\}}}u_{i}\left(\varphi^\pi_{i}\left(\left(\succeq_{i},\succeq'_{-i}\right),X\right)\right)\mbox{; and}\\
\max_{\succeq'_{-i}\in{\cal R}^{N\setminus \{i\}}}u_{i}\left(\varphi^\pi_{i}\left(\left(\succeq'_{i},\succeq'_{-i}\right),X\right)\right) &\leq \max_{\succeq'_{-i}\in{\cal R}^{N\setminus \{i\}}}u_{i}\left(\varphi^\pi_{i}\left(\left(\succeq_{i},\succeq'_{-i}\right),X\right)\right)
\end{align*}
for any additive utility function $u_i$ consistent with $\succeq_i$. It is clear that the first of these inequalities is implied by Theorem~\ref{thm:Maxminimizer}. The second one is easy to verify directly. The best-case scenario after $i$ reports truthfully is for her to receive her $k$ most-preferred objects, where $k\coloneqq \lvert\varphi^\pi_i(\cdot,X)\rvert$. This occurs when, for example, all other agents rank those $k$ objects at the bottom of their preference orders. As that is the best possible outcome for $i$, the second inequality follows.\footnote{We are grateful to Shiri Ron for pointing out the connection between maxmin strategy-proofness and non-obvious manipulability.}

\begin{cor}\label{cor:nonobvious}
Draft rules are not obviously manipulable.
\end{cor}

Taken together, Theorem~\ref{thm:Maxminimizer} and Corollary~\ref{cor:nonobvious} suggest that, although draft rules are not strategy-proof, they nevertheless enjoy some desirable incentive properties. First, when others' reported preferences are unknown, ambiguity-averse agents can ``insure'' themselves by reporting their preferences truthfully. Second, any profitable deviation (if one exists) is neither obvious nor easy to find.



\section{Unacceptable objects}
\label{s:unacc_ext}

In most sports drafts, upon picking a player, a team receives that player's contract rights. It can choose whether to sign him or her, trade the contract rights to another team, or simply retain the contract rights without offering a contract. In other words, contract rights are subject to free disposal. A pick's option value means that a team would never want to pass on their turn in the draft even if the remaining pool of players is below replacement level.

The implementation of the draft in the Australian Football League (AFL) is different: whenever a team chooses a player in one of the AFL drafts, it must sign that player to its squad subject to the applicable minimum wages. This violates free disposal and, as a result, teams frequently pass in later rounds of the AFL drafts as they find all remaining available players to be unacceptable.\footnote{For example, the Greater Western Sydney Giants passed in the last seven rounds of the nine-round 2016 AFL rookie draft.}

In this section, we extend the preference domain by allowing agents
to declare some objects unacceptable. The sets $N$ of
agents, $\mathbb{O}$ of potential objects, ${\cal X}$ of sets of
available objects and ${\cal B}$ of bundles are the same as in Section
\ref{Model}. Now a \emph{(strict) preference $\succeq_{i}$} for
agent $i$ is a complete, transitive, and antisymmetric binary relation
on $\mathbb{O}\cup\left\{ \omega\right\} $, where $\omega$ is a
``null object'' that represents receiving no real object. We write
$\succ_{i}$ for the asymmetric part of $\succeq_{i}$, i.e., for
all $x,y\in\mathbb{O}\cup\left\{ \omega\right\} $, $x\succ_{i}y$
if and only if $x\succeq_{i} y$ and $x\neq y$. Let ${\cal R}$ denote
the set of all strict preferences. The set of all preference profiles
$\succeq=\left(\succeq_{i}\right)_{i\in N}$ is denoted by ${\cal R}^{N}$.

Given $\succeq_{i}\in{\cal R}$ and $x\in\mathbb{O}\cup\left\{ \omega\right\} $,
$U\left(\succeq_{i},x\right)=\left\{ y\in\mathbb{O}\mid y\succeq_{i}x\right\} $
denotes the \emph{upper contour set of $x$ at $\succeq_{i}$.} We
say an object $x\in\mathbb{O}$ is \emph{acceptable} at $\succeq_{i}$
if $x\succ_{i}\omega$; $x$ is \emph{unacceptable} at $\succeq_{i}$
if $\omega\succ_{i}x$. We denote the set of acceptable objects at
$\succeq_{i}$ by $U\left(\succeq_{i}\right)\coloneqq U\left(\succeq_{i},\omega\right)$. Given a set $X\in{\cal X}$, let $\operatorname{top}_{\succeq_{i}}\left(X\right)$
denote the top-ranked alternative in $X\cup\{\omega\}$ according to $\succeq_{i}$.

The definition of the pairwise dominance extension must be adapted
to the present setting. Given a preference relation $\succeq_{i}\in{\cal R}$,
the \emph{pairwise dominance extension $\succeq_{i}^{PD}$} is the
binary relation on ${\cal B}$ defined as follows: for all $S,T\in{\cal B}$,
$S\succeq_{i}^{PD}T$ if and only if there is an injective function $\mu:T\cap U\left(\succeq_{i}\right)\to S\cap U\left(\succeq_{i}\right)$
such that $\mu\left(x\right)\succeq_{i}x$ for each $x\in T\cap U\left(\succeq_{i}\right)$.
Note that the relation $\succeq_{i}^{PD}$ ignores the presence of unacceptable objects in any compared bundles.\footnote{For example, if $a\succ_{i}\omega\succ_{i}b$, then $\left\{ a,b\right\} \succeq_{i}^{PD}\left\{ a\right\} $
	even though $\left\{ a,b\right\} $ contains an unacceptable object.
	Of course, we also have $\left\{ a\right\} \succeq_{i}^{PD}\left\{ a,b\right\}$,
	which implies that $\succeq_{i}^{PD}$ is \emph{not} antisymmetric
	(unlike in Section \ref{Model}).} This does not affect our results because, in what follows, we will
restrict attention to individually rational allocation rules.

\subsection{Properties of Allocation Rules}

This subsection introduces the properties used in the characterization with unacceptable objects. The first property is specific to the present setting. Given a problem $\left(\succeq,X\right)$, an allocation $A\in{\cal A}\left(X\right)$ is called \emph{individually rational at $\left(\succeq,X\right)$} if, for each agent $i\in N$, it holds that $A_{i}\subseteq U\left(\succeq_{i}\right)$.

\begin{property}
	\label{prop:IR}
	An allocation rule $\varphi$ is \emph{individually
		rational (IR)} if, for any problem $\left(\succeq,X\right)$, $\varphi\left(\succeq,X\right)$
	is individually rational at $\left(\succeq,X\right)$.
\end{property}

The size-based properties wRP and wEF1, as well as NW, must be adapted to account for acceptability. An allocation rule weakly respects priority $\pi$ if the assigned bundle of each agent contains no fewer objects that she finds acceptable than the bundle of any agent with lower priority according to~$\pi$.

\begin{rproperty}{prop:WRP}\label{prop:WRP*}
An allocation rule $\varphi$ is \emph{weakly respectful of the priority $\pi$ (wRP\textsuperscript{*}-$\pi$)} if, for any problem $\left(\succeq,X\right)$ and any agent $i\in N$,
\[
\lvert\varphi_{i}\left(\succeq,X\right)\cap U(\succeq_i)\rvert \geq\lvert\varphi_{j}\left(\succeq,X\right)\cap U(\succeq_i)\rvert \text{ for each }j\in N\text{ such that }i\mathrel{\pi}j.
\]
An allocation rule $\varphi$ is \emph{weakly respectful of a priority (wRP\textsuperscript{*})} if there exists a priority $\pi$ such that $\varphi$ satisfies wRP\textsuperscript{*}-$\pi$.
\end{rproperty}

Weak EF1 requires that, for any agents $i,j$, the number of agent~$i$'s acceptable objects assigned to agent~$j$ does not exceed the number assigned to $i$ by more than one.

\begin{rproperty}{prop:WEF1}\label{prop:WEF1*}
	An allocation rule $\varphi$ is \emph{weakly envy-free up to one object (wEF1\textsuperscript{*})} if, for any problem $\left(\succeq,X\right)$ and any agents $i,j\in N$,
	\[
	\lvert \varphi_{j}\left(\succeq,X\right)\cap U(\succeq_i)\rvert \leq  \lvert \varphi_i\left(\succeq,X\right)\cap U(\succeq_i)\rvert + 1.
	\]
\end{rproperty}


Given a preference profile $\succeq$, let $U\left(\succeq\right)=\bigcup_{i\in N}U\left(\succeq_{i}\right)$
denote the set of objects that are acceptable to some agent. An allocation
rule is non-wasteful if it always allocates available objects that
are acceptable to some agent.

\begin{rproperty}{prop:NW}
	\label{prop:NW*}
	An allocation rule $\varphi$ is \emph{non-wasteful
		(NW\textsuperscript{*})} if, for any problem $\left(\succeq,X\right)$, $U\left(\succeq\right)\cap X\subseteq\bigcup_{i\in N}\varphi_{i}\left(\succeq,X\right)$.
\end{rproperty}

In addition, RP-$\pi$, EF1, and RM are defined exactly as in Properties~\ref{prop:RP}, \ref{prop:EF1}, and \ref{prop:RM}, respectively, except with respect to the redefined pairwise-dominance notion. Furthermore, RP-$\pi$ and EF1 continue to imply their weaker counterparts, wRP\textsuperscript{*}-$\pi$ and wEF1\textsuperscript{*}. 

An additional invariance property is needed for our characterization
of draft rules. Given a preference relation $\succeq_{i}$, we say
$\succeq'_{i}$ is a \emph{truncation of $\succeq_{i}$} if there
exists $x\in U\left(\succeq_{i}\right)$ such that $U\left(\succeq'_{i}\right)=U\left(\succeq_{i},x\right)$
and $\succeq'_{i}\mid_{U\left(\succeq'_{i}\right)}=\succeq_{i}\mid_{U\left(\succeq'_{i}\right)}$.
Moreover, if $\succeq'_{i}$ is a \emph{truncation of $\succeq_{i}$}
then we say that $\succeq_{i}$ is an \emph{extension of $\succeq'_{i}$}.\footnote{For example, if $\mathbb{O}=\left\{ x_{1},x_{2},\dots\right\} $, $\succeq_{i}=x_{1},x_{2},x_{3},\omega,\dots$, and $\succeq'_{i}=x_{1},x_{2},\omega,\dots$, then $\succeq'_{i}$ is a truncation of $\succeq_{i}$ and $\succeq_{i}$ is an extension of $\succeq'_{i}$.} We call $\succeq_{i}$ the \emph{complete extension of $\succeq'_{i}$}
if $U\left(\succeq_{i}\right)=\mathbb{O}$ and $\succeq'_{i}\mid_{\mathbb{O}}=\succeq_{i}\mid_{\mathbb{O}}$.\footnote{Note that the complete extension of a preference relation is unique.}
A preference profile $\succeq=\left(\succeq_{i}\right)_{i\in N}$
is called the \emph{complete extension of $\succeq'=\left(\succeq'_{i}\right)_{i\in N}$
}if $\succeq_{i}$ is the complete extension of $\succeq'_{i}$ for
each $i\in N$. The following properties require that no agent can
profitably manipulate the outcome by reporting truncations or extensions.

\begin{property}
	\label{prop:TPEP}
	An allocation rule $\varphi$ is
	\begin{enumerate}
		\item[(A)] \emph{truncation-proof} \emph{(TP)} if, for any problem $\left(\succeq,X\right)$
		and any agent $i\in N$, $\varphi_{i}\left(\succeq,X\right)\succeq_{i}^{PD}\varphi_{i}\left(\left(\succeq'_{i},\succeq_{-i}\right),X\right)$
		whenever $\succeq'_{i}$ is a truncation of $\succeq_{i}$.
		\item[(B)] \emph{extension-proof (EP)} if, for any problem $\left(\succeq,X\right)$
		and any agent $i\in N$, $\varphi_{i}\left(\succeq,X\right)\succeq_{i}^{PD}\varphi_{i}\left(\left(\succeq'_{i},\succeq_{-i}\right),X\right)$
		whenever $\succeq'_{i}$ is an extension of $\succeq_{i}$.
	\end{enumerate}
\end{property}

It is clear that TP and EP are weak variants of strategy-proofness. If an agent is assigned
some bundle at a given problem, then \emph{truncation invariance}
requires that she is assigned the same bundle whenever she truncates
to some preference relation under which the original bundle is acceptable.

\begin{property}
	\label{prop:TI}
	An allocation rule $\varphi$ is \emph{truncation
		invariant (TI)} if, for any problem $\left(\succeq,X\right)$ and any
	agent $i\in N$, $\varphi_{i}\left(\succeq,X\right)=\varphi_{i}\left(\left(\succeq'_{i},\succeq_{-i}\right),X\right)$
	whenever $\succeq'_{i}$ is a truncation of $\succeq_{i}$ such that
	$\varphi_{i}\left(\succeq,X\right)\subseteq U\left(\succeq'_{i}\right)$.
\end{property}

The following proposition, which says that TI is implied by the conjunction
of IR, TP, and EP, highlights why TI is appealing.

\begin{prop}
	\label{Theorem 5-1}If $\varphi$ is an allocation rule satisfying IR, TP, and EP, then $\varphi$ satisfies TI.
\end{prop}

\subsection{Draft Rules}

The draft rules defined in this section are virtually identical to
those of Section \ref{Draft Rules}, except that they never allocate
any agent an unacceptable object. At each step of the draft procedure,
the relevant agent is assigned her most-preferred remaining object,
provided it is acceptable; if none of the remaining objects are acceptable,
then she is assigned the null object. Formally, the \emph{draft rule associated with $\pi$} is the allocation rule
$\varphi^{\pi}:{\cal R}^{N}\times{\cal X}\to{\cal A}$ that associates
with each problem $\left(\succeq,X\right)$ the allocation $\varphi^{\pi}\left(\succeq,X\right)\in{\cal A}\left(X\right)$
determined by the following procedure.

\paragraph*{\rule{1\columnwidth}{1pt}}

\paragraph{Algorithm 2:\label{par:Algorithm 2}}

\noun{U-Draft$\left(\succeq,X,\pi\right)$.}
\begin{elabeling}{00.00.0000}
	\item [{\emph{Input:}}] A problem $\left(\succeq,X\right)$ and a priority
	$\pi$.
	\item [{\emph{Output:}}] An allocation $\varphi^{\pi}\left(\succeq,X\right)\in{\cal A}\left(X\right)$.
\end{elabeling}
\begin{enumerate}
	\item Define the sequence $\left(s_{k}\right)_{k=1}^{K}$ of \emph{selections
	}recursively. Set:
	\begin{enumerate}
		\item $s_{1}=\operatorname{top}_{\succeq_{f^\pi \left(1\right)}}\left(X\right)$.
		\item for $k=2,\dots,K$, $s_{k}=\operatorname{top}_{\succeq_{f^\pi \left(k\right)}}\left(X\setminus\left\{ s_{1},\dots,s_{k-1}\right\} \right)$,
		where $K$ is the smallest integer such that $s_{K-\left(n-1\right)}=s_{K-\left(n-2\right)}=\cdots=s_{K-1}=s_{K}=\omega$.
	\end{enumerate}
	\item For each $i\in N$, set $\varphi_{i}^{\pi}\left(\succeq,X\right)=\left\{ s_{k} \mid f^\pi \left(k\right)=i,1\leq k\leq K\right\}\setminus\{\omega\} $.
	\item Return $\varphi^{\pi}\left(\succeq,X\right)=\left(\varphi_{i}^{\pi}\left(\succeq,X\right)\right)_{i\in N}$.
\end{enumerate}
\rule{1\columnwidth}{1pt}

Draft rules satisfy a number of appealing properties, even in the presence of unacceptable objects. The following theorem extends Theorem~\ref{thm:main characterization} by providing a characterization of the draft rule $\varphi^{\pi}$ in this environment.

\begin{thm}
	\label{thm:characterization extension}An allocation rule satisfies RP-$\pi$, EF1, NW\textsuperscript{*}, RM, IR, and TI if and only if it equals $\varphi^\pi$.
\end{thm}

The proof of Theorem~\ref{thm:characterization extension} establishes that RP-$\pi$ and EF1 can be replaced by wRP\textsuperscript{*}-$\pi$ and wEF1\textsuperscript{*}, respectively, in the statement of the theorem. The independence of the six properties is shown below.

\paragraph{Individually~rational~(IR).}

Given any priority $\pi$, let $\varphi$ denote the allocation rule
defined by $\varphi\left(\succeq,X\right)=\varphi^{\pi}\left(\succeq',X\right)$,
where $\succeq'$ is the complete extension of $\succeq$. Clearly,
$\varphi$ satisfies NW\textsuperscript{*}, RP-$\pi$, EF1, RM, and TI, but fails
IR.

\paragraph{Non-wasteful~(NW\textsuperscript{*}).}

Let $\varphi$ denote the null allocation rule, defined by $\varphi\left(\succeq,X\right)=\left(\emptyset\right)_{i\in N}$
for each problem $\left(\succeq,X\right)$. Then $\varphi$ satisfies
IR, RP-$\pi$, EF1, RM, and TI, but fails NW\textsuperscript{*}.

\paragraph{Weakly~respectful~of~a~priority~(wRP\textsuperscript{*}).}

It is easy to see that the snake draft (see Section~\ref{sss:main_char_result} for its definition) violates wRP\textsuperscript{*}, but it satisfies IR, NW\textsuperscript{*}, EF1, and RM. The arguments showing that the snake draft satisfies TI are identical to the analogous arguments in the proof of Theorem~\ref{thm:characterization extension}.

\paragraph{Weakly~envy-free~up~to~one~object~(wEF1\textsuperscript{*}).}

Given any priority $\pi$, let $\varphi$ denote the \emph{serial
	dictatorship}, recursively defined for any problem $\left(\succeq,X\right)$ by
\[
	\varphi_{\pi_1}\left(\succeq,X\right) =X\cap U\left(\succeq_{\pi_1}\right)~\text{and}~\varphi_{\pi_k}\left(\succeq,X\right) =\left(X\cap U\left(\succeq_{\pi_k}\right)\right)\setminus\bigcup_{\ell=1}^{k-1}\varphi_{\pi_\ell}\left(\succeq,X\right)
\]
for $k\in\{2,\ldots,n\}$. Then $\varphi$ satisfies IR, NW\textsuperscript{*}, RP-$\pi$, RM, and TI, but violates wEF1\textsuperscript{*}.

\paragraph{Resource monotonic~(RM).}

Consider the set $X'=\left\{ x_{1},\dots,x_{n+1}\right\} $ and a
preference profile $\succeq'$ such that
\begin{align*}
	\succeq'_{1}\mid_{X'} & =x_{1},x_{2},\dots,x_{n+1},\\
	\succeq'_{i}\mid_{X'} & =x_{2},x_{3},\dots,x_{n+1},x_{1}\text{ for all }i\in N\setminus\left\{ 1\right\} ,
\end{align*}
and $U\left(\succeq'_{i}\right)=X'$ for all $i\in N$. Let $\overline{{\cal R}^{N}}$
be the class of all preference profiles $\succeq$ such that, for
each $i\in N$, $\succeq_{i}$ is either a truncation or an extension
of $\succeq'_{i}$. Let $\pi$ denote the identity priority and consider the allocation rule $\varphi$ defined
by
\[
\varphi\left(\succeq,X\right)=\left(\left\{ x_{1}\right\} \cup\varphi_{1}^{\pi}\left(\succeq,X'\setminus\left\{ x_{1}\right\} \right),\varphi_{2}^{\pi}\left(\succeq,X'\setminus\left\{ x_{1}\right\} \right),\dots,\varphi_{n}^{\pi}\left(\succeq,X'\setminus\left\{ x_{1}\right\} \right)\right)
\]
whenever $X=X'$, $\succeq\in\overline{{\cal R}^{N}}$ and $x_{1}\succ_{1}\omega$,
and by $\varphi\left(\succeq,X\right)=\varphi^{\pi}\left(\succeq,X\right)$
otherwise.\footnote{That is, if $X=X'$, $\succeq\in\overline{{\cal R}^{N}}$ and $x_{1}\succ_{1}\omega$,
	then $\varphi$ returns the allocation that agrees with $\varphi^{\pi}\left(\succeq,X'\setminus\left\{ x_{1}\right\} \right)$,
	except that it augments agent 1's bundle with the object $x_{1}$.} Then $\varphi$ violates RM, since
\[
\varphi_{2}\left(\succeq',X'\setminus\left\{ x_{n+1}\right\} \right)=\left\{ x_{2}\right\} \succ_{2}^{\prime PD}\left\{ x_{3}\right\} =\varphi_{2}\left(\succeq',X'\right).
\]
It is straightforward to show that $\varphi$ satisfies NW\textsuperscript{*}, IR, RP-$\pi$,
EF1, and TI.

\paragraph{Truncation~invariant~(TI).}

Let $\pi$ denote the identity priority and let $\succeq'$ be a preference profile satisfying the following properties:
\begin{enumerate}
	\item $b\succ'_{1}x\succ'_{1}a\succ'_{1}\omega\text{ for all }x\in\mathbb{O}\setminus\left\{ a,b\right\} $.
	\item $a\succ'_{2}\omega\succ'_{2}b$.
	\item $U\left(\succeq'_{i}\right)=\emptyset$ for each $i\in N\setminus\{1,2\}$.
\end{enumerate}
Then define the allocation rule $\varphi$ by
\[
\varphi\left(\succeq,X\right)=\begin{cases}
	\left(X,\emptyset,\dots,\emptyset\right) & \text{if}~\succeq=\succeq'~\text{and}~X\subseteq\left\{ a,b\right\}; \\
	\varphi^{\pi}\left(\succeq,X\right) & \text{otherwise}.
\end{cases}
\]
It is easy to see that $\varphi$ satisfies IR, NW\textsuperscript{*}, RP-$\pi$, and
EF1.

To establish RM, it suffices to show that $\varphi_1(\succeq',X') \succeq^{\prime PD}_1 \varphi_1(\succeq',X)$ under the assumption that $X \subseteq \{a,b\}$ and $X \subseteq X' \nsubseteq \{a,b\}$. If $X = \{a\}$, then $x\in\varphi_1(\succeq',X')$ for some $x \in \mathbb{O} \setminus \{a\}$. If $X = \{b\}$, then $b\in\varphi_1(\succeq',X')$. If $X = \{a,b\}$, then $\{b,x\}\subseteq\varphi_1(\succeq',X')$ for some $x \in \mathbb{O} \setminus \{a,b\}$. In all cases it follows that $\varphi_1(\succeq',X') \succeq^{\prime PD}_1 \varphi_1(\succeq',X)$.

Finally, $\varphi$ fails TI, since if $\succeq'_{2}$ is a truncation
of $\succeq_{2}^{*}$ such that $a\succ_{2}^{*}b\succ_{2}^{*}\omega$,
then $\varphi_{2}\left(\left(\succeq_{2}^{*},\succeq'_{-2}\right),\left\{ a,b\right\} \right)=\left\{ a\right\} \subseteq U\left(\succeq'_{2}\right)$
but
\[
\varphi_{2}\left(\left(\succeq_{2}^{*},\succeq'_{-2}\right),\left\{ a,b\right\} \right)=\left\{ a\right\} \neq\emptyset=\varphi_{2}\left(\succeq',\left\{ a,b\right\} \right).
\]

\section{Conclusion}
\label{sec:conclusion}

In this paper, we present the first axiomatic characterization of the widely used draft mechanism, recast as a centralized allocation rule. The draft is the lone rule satisfying respect for priority, envy-freeness up to one object, non-wastefulness, and resource monotonicity. We argue that the first two properties are essential for competitive balance, an important desideratum in sports leagues where the draft is most prominently used. Despite lacking strategy-proofness, the draft retains some positive incentive properties: truth-telling is optimal for maxmin utility maximizers facing sufficient uncertainty about other agents' preference reports, and the draft is not obviously manipulable. Hence, the draft performs well in terms of efficiency, fairness, and incentive compatibility, three desiderata that are notoriously difficult to achieve simultaneously. Two impossibility results show that there does not exist a mechanism that can meaningfully improve on the draft's properties. Overall, the draft appears very well suited for its primary goal of redressing competitive imbalances in sports leagues.

\appendix

\renewcommand{\theequation}{\thesection.\arabic{equation}}
\setcounter{equation}{0}

\section{Proof Appendix}
\label{sec:appendix}

The following two lemmata are useful in proving Theorem~\ref{thm:main characterization}. Lemma~\ref{lem:RM lemma} gives a useful property of resource-monotonic allocation rules. It says that, if $\varphi$ is resource monotonic and agrees with $\varphi^{f}$ at $\left(\succeq,X\right)$ for some picking sequence $f$, then, whenever we add to the set $X$ of available objects an object $x$ that is worse for each agent than any object assigned to her at $\left(\succeq,X\right)$, the bundles assigned to the agents at the problems $\left(\succeq,X\right)$ and $\left(\succeq,X\cup\left\{ x\right\} \right)$ differ by at most $\left\{ x\right\} $ (and, thus, at most one agent receives a different bundle). Lemma~\ref{lem:bundle size lemma} shows that for any allocation chosen by a rule satisfying wRP-$\pi$ and wEF1, there is an agent who is critical in the sense that she is assigned the same number of objects as any agent ahead of her in the priority order and exactly one more object than any agent behind her in that order.

\begin{lem}
\label{lem:RM lemma}Suppose $\varphi$ is an allocation rule satisfying
RM. Consider a problem $\left(\succeq,X\right)$ such that $\varphi\left(\succeq,X\right)=\varphi^{f}\left(\succeq,X\right)$
for some picking sequence $f$. Let $x\in\mathbb{O}\setminus X$
satisfy $y\succ_{i}x$ for each $i\in N$ and each $y\in\varphi_{i}(\succeq,X)$.
Then $\varphi_{i}\left(\succeq,X\right)\subseteq\varphi_{i}\left(\succeq,X\cup\left\{ x\right\} \right)$
for each agent $i\in N$.
\end{lem}

\begin{proof}[\textbf{Proof of Lemma~\ref{lem:RM lemma}:}]Let $\left(s_{k}\right)_{k=1}^{\lvert X\rvert }$ be the sequence of
selections associated with the draft procedure \noun{Draft}$\left(\succeq,X,f\right)$.
We show by induction that $s_{k}\in\varphi_{f\left(k\right)}\left(\succeq,X\cup\left\{ x\right\} \right)$
for each $k=1,\dots,\lvert X\rvert $.

\textbf{Base case $\left(k=1\right)$.} Since $s_{1}=\operatorname{top}_{\succeq_{f\left(1\right)}}\left(X\right)$,
RM implies that $s_{1}\in\varphi_{f\left(1\right)}\left(\succeq,X\cup\left\{ x\right\} \right)$.

\textbf{Inductive step.} Suppose that, for some $k\in\left\{ 2,\dots,\lvert X\rvert \right\} $,
$s_{\ell}\in\varphi_{f\left(\ell\right)}\left(\succeq,X\cup\left\{ x\right\} \right)$
whenever $1\leq\ell<k$. We must show that $s_{k}\in\varphi_{f\left(k\right)}\left(\succeq,X\cup\left\{ x\right\} \right)$.

Observe that the induction hypothesis implies that $\varphi_{i}\left(\succeq,X\right)\cap\left\{ s_{1},\dots,s_{k-1}\right\} =\varphi_{i}\left(\succeq,X\cup\left\{ x\right\} \right)\cap\left\{ s_{1},\dots,s_{k-1}\right\} $
for each $i\in N$. Since $s_{k}\in\varphi_{f\left(k\right)}\left(\succeq,X\right)$
and $s_{k}=\operatorname{top}_{\succeq_{f\left(k\right)}}\left(X\setminus\left\{ s_{1},\dots,s_{k-1}\right\} \right)$,
RM implies that $s_{k}\in\varphi_{f\left(k\right)}\left(\succeq,X\cup\left\{ x\right\} \right)$,
as desired.

It follows that $s_{k}\in\varphi_{f\left(k\right)}\left(\succeq,X\cup\left\{ x\right\} \right)$
for each $k=1,\dots,\lvert X\rvert $. Hence, $\varphi_{i}\left(\succeq,X\right)\subseteq\varphi_{i}\left(\succeq,X\cup\left\{ x\right\} \right)$
for each $i\in N$.
\end{proof}

\begin{lem}
\label{lem:bundle size lemma}Suppose $\varphi$ is an allocation
rule satisfying wRP-$\pi$ and wEF1. Then, for each problem $\left(\succeq,X\right)$, there is some agent $i^* \in N$ such that
\begin{align*}
\lvert \varphi_{j}\left(\succeq,X\right)\rvert  & =\lvert \varphi_{i^*}\left(\succeq,X\right)\rvert \text{ whenever } j\mathrel{\pi} i^*\\
\text{and}\quad\lvert \varphi_{j}\left(\succeq,X\right)\rvert  & =\lvert \varphi_{i^*}\left(\succeq,X\right)\rvert -1\text{ whenever }i^*\mathrel{\pi} j.
\end{align*}
\end{lem}

\begin{proof}[\textbf{Proof of Lemma~\ref{lem:bundle size lemma}:}] Consider any problem $\left(\succeq,X\right)$ and, without loss of generality, assume $\pi$ to be the identity priority. By the definition of wEF1, the sizes of any two agents' bundles $\lvert \varphi_{i}\left(\succeq,X\right)\rvert$ and $\lvert \varphi_{j}\left(\succeq,X\right)\rvert$ differ by at most one. Moreover, by wRP-$\pi$, the agents' bundle sizes $\lvert \varphi_{j}\left(\succeq,X\right)\rvert $ are nonincreasing in $j$. The desired conclusion follows directly from these two observations.
\end{proof}

\begin{proof}[\textbf{Proof of Theorem~\ref{thm:main characterization}:}]

Without loss of generality, for the duration of this proof we assume that $\pi$ is the identity priority. We start with the \textbf{``if''} direction. It is clear that $\varphi^{\pi}$ satisfies NW and RP-$\pi$.

\textbf{(EF1)} To see that $\varphi^{\pi}$ satisfies EF1, consider
any problem $\left(\succeq,X\right)$ and distinct agents $i,j\in N$
with $i<j$. It suffices to show that there exists $S\subseteq\varphi_{i}^{\pi}\left(\succeq,X\right)$
such that $\varphi_{j}^{\pi}\left(\succeq,X\right)\succeq_{j}^{PD}\varphi_{i}^{\pi}\left(\succeq,X\right)\setminus S$.
If $\varphi_{j}^{\pi}\left(\succeq,X\right)=\emptyset$, then the
definition of $\varphi^{\pi}$ implies that $\lvert \varphi_{i}^{\pi}\left(\succeq,X\right)\rvert \leq1$;
hence, $S=\varphi_{i}^{\pi}\left(\succeq,X\right)$ will do. So assume
$\varphi_{j}^{\pi}\left(\succeq,X\right)\neq\emptyset$. Then we may
write $\varphi_{i}^{\pi}\left(\succeq,X\right)=\left\{ x_{1}\succ_{i}\cdots\succ_{i}x_{k}\right\} $
and $\varphi_{j}^{\pi}\left(\succeq,X\right)=\left\{ y_{1}\succ_{j}\cdots\succ_{j}y_{\ell}\right\} $
for some $\ell\geq1$ and $k\in\left\{ \ell,\ell+1\right\} $ (because
$i<j$). In the draft procedure \noun{Draft}$\left(\succeq,X,f^{\pi}\right)$,
for each $t=2,\dots,k$, object $y_{t-1}$ was assigned to agent $j$
before object $x_{t}$ was assigned to agent $i$; hence, $y_{t-1}\succ_{j}x_{t}$
for each $t=2,\dots,k$. Consequently, $\varphi_{j}\left(\succeq,X\right)\succeq_{j}^{PD}\varphi_{i}\left(\succeq,X\right)\setminus\left\{ x_{1}\right\} $
and EF1 holds.

\textbf{(RM)} By transitivity of $\succeq_{i}^{PD}$ it suffices to
show that, given a problem $\left(\succeq,X\right)$ and an object
$x\in\mathbb{O}\setminus X$, one has $\varphi_{i}^{\pi}\left(\succeq,X\cup\left\{ x\right\} \right)\succeq_{i}^{PD}\varphi_{i}^{\pi}\left(\succeq,X\right)$
for each $i\in N$. Suppose $\left(s_{k}\right)_{k=1}^{\lvert X\rvert }$
is the sequence of selections associated with \noun{Draft}$\left(\succeq,X,f^{\pi}\right)$,
and $\left(s'_{k}\right)_{k=1}^{\lvert X\rvert +1}$ is the sequence
of selections associated with \noun{Draft}$\left(\succeq,X\cup\left\{ x\right\} ,f^{\pi}\right)$.
We argue by induction that $X\cup\left\{ x\right\} \setminus\left\{ s'_{1},\dots,s'_{k}\right\} \supseteq X\setminus\left\{ s_{1},\dots,s_{k}\right\} $
for each $k=0,1,\dots,\lvert X\rvert -1$.

\textbf{Base case $\left(k=0\right)$. }Clearly, $X\cup\left\{ x\right\} \setminus\emptyset\supseteq X\setminus\emptyset$.

\textbf{Inductive step. }Suppose $X\cup\left\{ x\right\} \setminus\left\{ s'_{1},\dots,s'_{k-1}\right\} \supseteq X\setminus\left\{ s_{1},\dots,s_{k-1}\right\} $
for some $k\in\left\{ 1,\dots,\lvert X\rvert -1\right\} $. We must
show that $X\cup\left\{ x\right\} \setminus\left\{ s'_{1},\dots,s'_{k}\right\} \supseteq X\setminus\left\{ s_{1},\dots,s_{k}\right\} $.
Let $y\in X\setminus\left\{ s_{1},\dots,s_{k}\right\} $. The induction
hypothesis implies that $X\setminus\left\{ s_{1},\dots,s_{k}\right\} \subseteq X\setminus\left\{ s_{1},\dots,s_{k-1}\right\} \subseteq X\cup\left\{ x\right\} \setminus\left\{ s'_{1},\dots,s'_{k-1}\right\} $,
which means that $y\in X\cup\left\{ x\right\} \setminus\left\{ s'_{1},\dots,s'_{k-1}\right\} $.
Let $i$ be the agent who is assigned an object at step $k$ of the
draft procedure; that is, $i=f^{\pi}\left(k\right)$. Then
\[
s'_{k}=\operatorname{top}_{\succeq_{i}}\left(X\cup\left\{ x\right\} \setminus\left\{ s'_{1},\dots,s'_{k-1}\right\} \right)\succeq_{i}\operatorname{top}_{\succeq_{i}}\left(X\setminus\left\{ s_{1},\dots,s_{k-1}\right\} \right)=s_{k}\succ_{i}y,
\]
which means that $s'_{k}\succ_{i}y$. Hence, $y\in X\cup\left\{ x\right\} \setminus\left\{ s'_{1},\dots,s'_{k}\right\} $,
as we needed to show.

Since $X\cup\left\{ x\right\} \setminus\left\{ s'_{1},\dots,s'_{k}\right\} \supseteq X\setminus\left\{ s_{1},\dots,s_{k}\right\} $
for each $k=0,1,\dots,\lvert X\rvert -1$, we have $s'_{k}\succeq_{f^{\pi}\left(k\right)}s_{k}$
for each $k=1,\dots,\lvert X\rvert $. Therefore, RM holds.

For the \textbf{``only if''} direction, assume that $\varphi$ satisfies wRP-$\pi$, wEF1, NW, and RM. Start by letting
$f^{\pi}:\mathbb{N}\to N$ denote the picking sequence associated
with $\pi$, defined by $\left(f^{\pi}\left(k\right)\right)_{k\in\mathbb{N}}=\left(1,2,\dots,n,1,2,\dots,n,\dots\right)$.
Given any problem $\left(\succeq,X\right)$, we must show that $\varphi\left(\succeq,X\right)=\varphi^{\pi}\left(\succeq,X\right)$.

Let $\left(s_{k}\right)_{k=1}^{\lvert X\rvert }$ denote the sequence
of selections associated with the draft procedure \noun{Draft}$\left(\succeq,X,f^{\pi}\right)$.
For each $k=1,\dots,\lvert X\rvert $, let $S_{k}\coloneqq\left\{ s_{1},\dots,s_{k}\right\} $.
We argue by induction that $\varphi\left(\succeq,S_{k}\right)=\varphi^{\pi}\left(\succeq,S_{k}\right)$
for each $k=1,\dots,\lvert X\rvert $.

\textbf{Base case $\left(k=1\right)$.} Since $\varphi$ satisfies
NW and wRP-$\pi$, agent 1 must receive object $s_{1}$ at the problem
$\left(\succeq,S_{1}\right)$. Hence, $\varphi\left(\succeq,S_{1}\right)=\varphi^{\pi}\left(\succeq,S_{1}\right)$.

\textbf{Inductive step.} Suppose that $\varphi\left(\succeq,S_{k-1}\right)=\varphi^{\pi}\left(\succeq,S_{k-1}\right)$
for some $k\in\left\{ 2,\dots,\lvert X\rvert \right\} $. We must show
that $\varphi\left(\succeq,S_{k}\right)=\varphi^{\pi}\left(\succeq,S_{k}\right)$.

\paragraph{Step 1: Objects in $S_{k-1}$ are assigned to the \textquotedblleft correct
agents\textquotedblright{} at $\left(\succeq,S_{k}\right)$, i.e.,
$s_{\ell}\in\varphi_{f^{\pi}\left(\ell\right)}\left(\succeq,S_{k}\right)$
for each $\ell=1,\dots,k-1$.}

Observe that $s_{k}\in\mathbb{O}\setminus S_{k-1}$ is an object
that is worse for each agent than any object assigned to her at $\left(\succeq,S_{k-1}\right)$,
i.e., 
\[
\text{for each }j\in N,\quad\left(y\succ_{j}s_{k}\text{ for each }y\in\varphi_{j}\left(\succeq,S_{k-1}\right)\right).
\]
Therefore, Lemma~\ref{lem:RM lemma} and the inductive hypothesis
imply that $\varphi_{j}\left(\succeq,S_{k-1}\right)\subseteq\varphi_{j}\left(\succeq,S_{k}\right)$
for each agent $j\in N$. The inductive hypothesis implies $\varphi\left(\succeq,S_{k-1}\right)=\varphi^{\pi}\left(\succeq,S_{k-1}\right)$,
and therefore $s_{\ell}\in\varphi_{f^{\pi}\left(\ell\right)}\left(\succeq,S_{k-1}\right)\subseteq\varphi_{f^{\pi}\left(\ell\right)}\left(\succeq,S_{k}\right)$
for each $\ell=1,\dots,k-1$, as claimed.

\paragraph{Step 2: Object $s_{k}$ is assigned to agent $f^{\pi}\left(k\right)$,
i.e., $s_{k}\in\varphi_{f^{\pi}\left(k\right)}\left(\succeq,S_{k}\right)$.}

Since $\varphi\left(\succeq,S_{k-1}\right)=\varphi^{\pi}\left(\succeq,S_{k-1}\right)$,
the definition of \noun{Draft}$\left(\succeq,S_{k-1},f^{\pi}\right)$
implies that $f^{\pi}\left(k-1\right)$ is an agent such that
\begin{align*}
\lvert \varphi_{j}\left(\succeq,S_{k-1}\right)\rvert  & =\lvert \varphi_{f^{\pi}\left(k-1\right)}\left(\succeq,S_{k-1}\right)\rvert \text{ whenever }j\leq f^{\pi}\left(k-1\right)\\
\text{and }\lvert\varphi_{j}\left(\succeq,S_{k-1}\right)\rvert  & =\lvert \varphi_{f^{\pi}\left(k-1\right)}\left(\succeq,S_{k-1}\right)\rvert -1\text{ whenever }j>f^{\pi}\left(k-1\right).
\end{align*}
By Step 1 and NW, it holds that $\varphi_{i}\left(\succeq,S_{k}\right)=\varphi_{i}\left(\succeq,S_{k-1}\right)\cup\left\{ s_{k}\right\} $
for exactly one agent $i\in N$, and $\varphi_{j}\left(\succeq,S_{k-1}\right)=\varphi_{j}\left(\succeq,S_{k}\right)$
for each agent $j\in N\setminus\left\{ i\right\} $. We must show
that $i=f^{\pi}\left(k\right)$.

\begin{casenv}
\item If $i>f^{\pi}\left(k\right)$, i.e., $s_{k}$ is allocated to some
agent with lower priority than $f^{\pi}\left(k\right)$, then the
preceding discussion implies that
\[
\lvert \varphi_{f^{\pi}\left(k\right)}\left(\succeq,S_{k}\right)\rvert =\lvert \varphi_{f^{\pi}\left(k\right)}\left(\succeq,S_{k-1}\right)\rvert =\lvert \varphi_{i}\left(\succeq,S_{k-1}\right)\rvert =\lvert \varphi_{i}\left(\succeq,S_{k}\right)\rvert -1,
\]
which contradicts wRP-$\pi$.
\item If $i<f^{\pi}\left(k\right)$, i.e., $s_{k}$ is allocated to some
agent with higher priority than $f^{\pi}\left(k\right)$, then the
preceding discussion shows that
\[
\lvert \varphi_{f^{\pi}\left(k\right)}\left(\succeq,S_{k}\right)\rvert =\lvert \varphi_{f^{\pi}\left(k\right)}\left(\succeq,S_{k-1}\right)\rvert =\lvert \varphi_{i}\left(\succeq,S_{k-1}\right)\rvert -1=\lvert \varphi_{i}\left(\succeq,S_{k}\right)\rvert -2,
\]
which contradicts wEF1.
\end{casenv}

Putting this all together, we have that $i=f^{\pi}\left(k\right)$
and $s_{k}\in\varphi_{f^{\pi}\left(k\right)}\left(\succeq,S_{k}\right)$.

By the principle of induction, we have $\varphi\left(\succeq,S_{k}\right)=\varphi^{\pi}\left(\succeq,S_{k}\right)$
for each $k=1,\dots,\lvert X\rvert $. In particular, $\varphi\left(\succeq,X\right)=\varphi\left(\succeq,S_{\lvert X\rvert }\right)=\varphi^{\pi}\left(\succeq,S_{\lvert X\rvert }\right)=\varphi^{\pi}\left(\succeq,X\right)$,
as we needed to show.
\end{proof}

\begin{proof}[\textbf{Proof of Theorem~\ref{thm:impossibility 1}:}] Without loss of generality, assume that $\pi$ is the identity priority.
Consider any set $X\in{\cal X}$ with $\lvert X\rvert =n+1$, say $X=\left\{ x_{1},\dots,x_{n+1}\right\} $,
and let $\succeq$ be a preference profile such that
\begin{align*}
\succeq_{1}\mid_{X} & =x_{1},x_{2},\dots,x_{n+1}\\
\text{and }\succeq_{i}\mid_{X} & =x_{2},\dots,x_{n+1},x_{1},\text{ for all }i\in N\setminus\left\{ 1\right\} .
\end{align*}
Toward contradiction, assume that $\varphi$ satisfies RP-$\pi$,
EF1, NW, and wSP, and consider the allocation $\varphi\left(\succeq,X\right)$.
Lemma~\ref{lem:bundle size lemma}, RP-$\pi$, EF1, and NW imply that agent
1 receives two objects while every agent $i\in N\setminus\left\{ 1\right\}$
receives one; moreover, RP-$\pi$ implies that agent 1 receives
her best object. Consequently, 
\[
\varphi_{1}\left(\succeq,X\right)\in\left\{ \left\{ x_{1},x_{2}\right\} ,\left\{ x_{1},x_{3}\right\} ,\dots,\left\{ x_{1},x_{n+1}\right\} \right\} .
\]
Suppose $\succeq'_{1}$ is a preference relation that agrees with
$\succeq_{1}$ on $X$ except that it interchanges the ranking of
objects $x_{1}$ and $x_{2}$; that is, let
\[
\succeq'_{1}\mid_{X}=x_{2},x_{1},x_{3},\dots,x_{n+1}.
\]
Consider the allocation $\varphi\left(\left(\succeq'_{1},\succeq_{-1}\right),X\right)$. Lemma~\ref{lem:bundle size lemma}, RP-$\pi$, EF1, and NW imply that agent 1 receives two objects and each other agent receives one, while RP-$\pi$ implies that 1 receives $x_2$. Additionally, EF1 implies that $\varphi_{1}\left(\left(\succeq'_{1},\succeq_{-1}\right),X\right)=\left\{ x_{1},x_{2}\right\} $ as any other two-object bundle containing $x_2$ would violate EF1 for agent 1 and the agent receiving the bundle $\{x_1\}$. By wSP, $\left\{ x_{1},x_{2}\right\} =\varphi_{1}\left(\left(\succeq'_{1},\succeq_{-1}\right),X\right)\nsucc_{1}^{PD}\varphi_{1}\left(\succeq,X\right)$,
which implies that
\[
\varphi_{1}\left(\succeq,X\right)=\varphi_{1}\left(\left(\succeq'_{1},\succeq_{-1}\right),X\right)=\left\{ x_{1},x_{2}\right\} .
\]
It follows from RP-$\pi$, NW, and EF1 that
\[
\varphi_{i}\left(\succeq,X\right)=\varphi_{i}\left(\left(\succeq'_{1},\succeq_{-1}\right),X\right)=\left\{ x_{i+1}\right\} \text{ for all }i\in N\setminus\left\{ 1\right\} .
\]
Now suppose that $\succeq'_{n}$ is a preference relation that agrees
with $\succeq_{n}$ on $X$ except that it interchanges the ranking
of objects $x_{1}$ and $x_{n+1}$; that is, let
\[
\succeq'_{n}\mid_{X}=x_{2},\dots,x_{n},x_{1},x_{n+1}.
\]

Consider the allocation $\varphi\left(\left(\succeq'_{n},\succeq_{-n}\right),X\right)$. As above, Lemma~\ref{lem:bundle size lemma}, RP-$\pi$, EF1, and NW imply that agent 1 receives two objects, one of which is her best object $x_1$, while each other agent receives one object.

If $\varphi_{1}\left(\left(\succeq'_{n},\succeq_{-n}\right),X\right)=\left\{ x_{1},x_{k}\right\}$
for some $k\in\left\{ 2,\dots,n\right\} $, then EF1 implies that $\varphi_{n}\left(\left(\succeq'_{n},\succeq_{-n}\right),X\right)\neq\{x_{n+1}\}$.
If $n=2$, this violates NW. If $n\geq 3$, it follows that there is an agent $i\in N\setminus\left\{1,n\right\}$ for whom $\varphi_{i}\left(\left(\succeq'_{n},\succeq_{-n}\right),X\right)=\left\{x_{n+1}\right\}$. This violates RP-$\pi$ as $i$ envies $n$. Therefore, $\varphi_{1}\left(\left(\succeq'_{n},\succeq_{-n}\right),X\right)=\left\{ x_{1},x_{n+1}\right\}$.

Analogously, if $\varphi_{n}\left(\left(\succeq'_{n},\succeq_{-n}\right),X\right)\neq\left\{ x_{n}\right\} $ (which is possible only when $n\geq 3$ due to NW), then $\varphi_{i}\left(\left(\succeq'_{n},\succeq_{-n}\right),X\right)=\left\{ x_{n}\right\}$ for some $i\in N\setminus\left\{ 1,n\right\}$ and agent $i$ envies agent $n$, thereby violating RP-$\pi$. Thus, $\varphi_{n}\left(\left(\succeq'_{n},\succeq_{-n}\right),X\right)=\left\{ x_{n}\right\}$. But then $\left\{ x_{n}\right\} =\varphi_{n}\left(\left(\succeq'_{n},\succeq_{-n}\right),X\right)\succ_{n}^{PD}\varphi_{n}\left(\succeq,X\right)=\left\{ x_{n+1}\right\} $,
which violates wSP!
\end{proof}

\begin{proof}[\textbf{Proof of Theorem~\ref{thm:impossibility 2}:}] Toward contradiction, suppose $\varphi$ satisfies EFF, EF1, and wSP.

First consider the case with $n=2$ agents. Given $X=\left\{ a,b,c,d\right\} \subseteq\mathbb{O}$,
let $\succeq$ and $\succeq'$ be preference profiles such that
\begin{align*}
\succeq_{1} & =a,b,c,d,\dots; &  &  & \succeq'_{1} & =b,d,a,c,\dots;\\
\succeq_{2} & =b,a,d,c,\dots; &  &  & \succeq'_{2} & =d,b,c,a,\dots.
\end{align*}
Then EFF and EF1 imply that
\[
\varphi\left(\succeq,X\right)=\left(\left\{ a,c\right\} ,\left\{ b,d\right\} \right)\quad\text{and}\quad\varphi\left(\succeq',X\right)=\left(\left\{ a,b\right\} ,\left\{ c,d\right\} \right).
\]
By wSP for agent 2, we have $\{b, d\} = \varphi_2\left( \succeq , X \right) \nsucc^{\prime PD}_2 \varphi_2\left( \left( \succeq_1, \succeq'_2 \right), X \right)$, which, together with EFF and EF1, implies $\varphi\left(\left(\succeq_{1},\succeq'_{2}\right),X\right)=\left(\left\{ a,c\right\} ,\left\{ b,d\right\} \right)$.
But then 
\[
\varphi_{1}\left(\succeq',X\right)=\left\{ a,b\right\} \succ_{1}^{PD}\left\{ a,c\right\} =\varphi_{1}\left(\left(\succeq_{1},\succeq'_{2}\right),X\right),
\]
which violates wSP for agent 1!

Using EFF, one can extend the above argument to the case with $n\geq3$
agents by adding $2\left(n-2\right)$ objects and $n-2$ agents who bottom-rank the objects
in $\left\{ a,b,c,d\right\} $.
\end{proof}

\begin{proof}[\textbf{Proof of Theorem~\ref{thm:Maxminimizer}:}] Without loss of generality, let $\pi$ denote the identity priority.
If $\lvert X\rvert <i$, then agent $i$ does not receive an object
at any preference profile and we are done. Assuming $\lvert X\rvert \geq i$,
let $k\in\mathbb{N}$ be such that $\left(k-1\right)n+i\leq\lvert X\rvert <kn+i$,
so that agent $i$ receives a bundle with $k$ objects at each preference profile.

Fix some $\succeq_{i}\in{\cal R}$ and suppose $u_{i}$ is an additive
utility function consistent with $\succeq_{i}$. Let $\succeq_{-i}\in{\cal R}^{N\setminus \{i\}}$
denote the profile such that $\succeq_{j}=\succeq_{i}$ for each $j\in N\setminus \{i\}$.
Clearly, $\succeq_{-i}$ solves the inner minimization problem at
$\succeq'_{i}=\succeq_{i}$, i.e.,
\[
\min_{\succeq'_{-i}\in{\cal R}^{N\setminus \{i\}}}u_{i}\left(\varphi_{i}^{\pi}\left(\left(\succeq_{i},\succeq'_{-i}\right),X\right)\right)=u_{i}\left(\varphi_{i}^{\pi}\left(\left(\succeq_{i},\succeq_{-i}\right),X\right)\right).
\]
It follows that
\[
\max_{\succeq'_{i}\in{\cal R}}\left[\min_{\succeq'_{-i}\in{\cal R}^{N\setminus \{i\}}}u_{i}\left(\varphi_{i}^{\pi}\left(\left(\succeq'_{i},\succeq'_{-i}\right),X\right)\right)\right]\geq u_{i}\left(\varphi_{i}^{\pi}\left(\left(\succeq_{i},\succeq_{-i}\right),X\right)\right).
\]
For the reverse inequality, it suffices to show that, for any $\succeq'_{i}\in{\cal R}$,
one has
\[
u_{i}\left(\varphi_{i}^{\pi}\left(\left(\succeq_{i},\succeq_{-i}\right),X\right)\right)\geq\min_{\succeq'_{-i}\in{\cal R}^{N\setminus \{i\}}}u_{i}\left(\varphi_{i}^{\pi}\left(\left(\succeq'_{i},\succeq'_{-i}\right),X\right)\right).
\]
To this end, it is enough to show that, for any $\succeq'_{i}\in{\cal R}$,
\begin{equation}
\varphi_{i}^{\pi}\left(\left(\succeq_{i},\succeq_{-i}\right),X\right)\succeq_{i}^{PD}\varphi_{i}^{\pi}\left(\left(\succeq'_{i},\succeq_{-i}\right),X\right).\label{eq:5-1}
\end{equation}
Let $\succeq'_{i}\in{\cal R}$, $\succeq \coloneqq (\succeq_i, \succeq_{-i})$, and $\succeq' \coloneqq (\succeq'_i, \succeq_{-i})$, and assume
that
\[
\varphi_{i}^{\pi}\left(\succeq,X\right)=\left\{ x_{1}\succ_{i}\cdots\succ_{i}x_{k}\right\} \quad\text{and}\quad\varphi_{i}^{\pi}\left(\succeq',X\right)=\left\{ y_{1}\succ_{i}\cdots\succ_{i}y_{k}\right\} .
\]

For each $\ell \in \{1, \dots, k\}$, denote $X_\ell \coloneqq \{x \in X \mid x \succ_i x_\ell \}$.

\begin{claim}\label{CountingClaim}
    For each $\ell \in \{1,\dots,k\}$, agent~$i$ receives at most $\ell - 1$ objects from $X_\ell$ in problem $(\succeq',X)$: $$|\varphi^\pi_i(\succeq',X) \cap X_\ell| \leq \ell - 1.$$
\end{claim}

\begin{subproof}
    Fix $\ell \in \{1, \dots, k\}$. Then $X_\ell$ contains exactly $|X_\ell| = (\ell - 1)|N| + (i - 1)$ objects.
    
    Consider the step of \noun{Draft}$(\succeq',X, f^\pi)$ at which agent~$i$ is assigned her $\ell$th object, i.e., step $t = (\ell - 1) |N| + i$. Before step~$t$, agents in $N \setminus \{i\}$ were already assigned 
    $$t - \ell = (\ell - 1) |N| + i -\ell = |X_\ell| - (\ell - 1)$$ objects. Thus, at the beginning of every step $s < t$ there were objects in $X_\ell$ remaining, and since all agents in $N \setminus \{i\}$ share the same preference $\succeq_i$ at $\succeq'$, these agents must have been assigned only objects from $X_\ell$. Consequently, at step~$t$, the number of objects in $X_\ell$ that were not already assigned to agents in $N \setminus \{i\}$ is
    $$
    |X_\ell| - (t - \ell) = \ell - 1.
    $$
    Thus, agent~$i$ receives at most $\ell - 1$ objects from $X_\ell$.
\end{subproof}

To complete the proof of (\ref{eq:5-1}), we show that $x_{\ell}\succeq_{i}y_{\ell}$
for each $\ell \in \{1,\dots,k\}$. Suppose, for a contradiction, that $y_\ell \succ_i x_\ell$ for some $\ell \in \{1, \dots, k\}$. Then, since $y_1 \succ_i \cdots \succ_i y_\ell \succ_i x_\ell$, we must have $\{y_1, \dots, y_\ell \} \subseteq \varphi^\pi_i(\succeq', X) \cap X_\ell$. Consequently, $|\varphi^\pi_i(\succeq',X) \cap X_\ell | \geq \ell$, contradicting Claim~\ref{CountingClaim}. Hence, (\ref{eq:5-1}) holds.
\end{proof}

\begin{proof}[\textbf{Proof of Proposition~\ref{Theorem 5-1}:}] Suppose $\varphi$ satisfies IR, TP, and EP. Given any problem $\left(\succeq,X\right)$
and any agent $i\in N$, let $\succeq'_{i}$ be a truncation of $\succeq_{i}$
such that $\varphi_{i}\left(\succeq,X\right)\subseteq U\left(\succeq'_{i}\right)$.
By TP, we have $\varphi_{i}\left(\succeq,X\right)\succeq_{i}^{PD}\varphi_{i}\left(\left(\succeq'_{i},\succeq_{-i}\right),X\right)$,
which means that 
\[
\lvert \varphi_{i}\left(\succeq,X\right)\cap U\left(\succeq_{i}\right)\rvert \geq\lvert \varphi_{i}\left(\left(\succeq'_{i},\succeq_{-i}\right),X\right)\cap U\left(\succeq_{i}\right)\rvert .
\]
On the other hand, EP implies that $\varphi_{i}\left(\left(\succeq'_{i},\succeq_{-i}\right),X\right)\succeq_{i}^{\prime PD}\varphi_{i}\left(\succeq,X\right)$,
which means that
\[
\lvert \varphi_{i}\left(\left(\succeq'_{i},\succeq_{-i}\right),X\right)\cap U\left(\succeq'_{i}\right)\rvert \geq\lvert \varphi_{i}\left(\succeq,X\right)\cap U\left(\succeq'_{i}\right)\rvert .
\]
It follows from IR and our hypothesis about $\succeq'_{i}$ that $\varphi_{i}\left(\left(\succeq'_{i},\succeq_{-i}\right),X\right)\subseteq U\left(\succeq'_{i}\right)\subseteq U\left(\succeq_{i}\right)$
and $\varphi_{i}\left(\succeq,X\right)\subseteq U\left(\succeq'_{i}\right)\subseteq U\left(\succeq_{i}\right)$.
Consequently, 
\begin{equation}
\lvert \varphi_{i}\left(\left(\succeq'_{i},\succeq_{-i}\right),X\right)\rvert =\lvert \varphi_{i}\left(\succeq,X\right)\rvert .\label{eq:7-3-1}
\end{equation}
If $\varphi_{i}\left(\succeq,X\right)=\emptyset$, then (\ref{eq:7-3-1})
implies that $\varphi_{i}\left(\left(\succeq'_{i},\succeq_{-i}\right),X\right)=\emptyset$
and we're done. So we may assume that 
\[
\varphi_{i}\left(\succeq,X\right)=\left\{ x_{1}\succ_{i}\cdots\succ_{i}x_{k}\right\} \quad\text{and}\quad\varphi_{i}\left(\left(\succeq'_{i},\succeq_{-i}\right),X\right)=\left\{ y_{1}\succ_{i}\cdots\succ_{i}y_{k}\right\} 
\]
for some $k\geq1$.

Since $\varphi_{i}\left(\succeq,X\right)\succeq_{i}^{PD}\varphi_{i}\left(\left(\succeq'_{i},\succeq_{-i}\right),X\right)$
and $\succeq_{i}$ agrees with $\succeq'_{i}$ on $\varphi_{i}\left(\succeq,X\right)\cup\varphi_{i}\left(\left(\succeq'_{i},\succeq_{-i}\right),X\right)\subseteq U\left(\succeq'_{i}\right)$,
we have that $x_{\ell}\succeq_{i}y_{\ell}$ for each $\ell=1,\dots,k$.
By EP, also $\varphi_{i}\left(\left(\succeq'_{i},\succeq_{-i}\right),X\right)\succeq_{i}^{\prime PD}\varphi_{i}\left(\succeq,X\right)$,
which means that $y_{\ell}\succeq_{i}x_{\ell}$ for each $\ell=1,\dots,k$.
It follows that $x_{\ell}=y_{\ell}$ for each $\ell=1,\dots,k$. That
is, $\varphi_{i}\left(\succeq,X\right)=\varphi_{i}\left(\left(\succeq'_{i},\succeq_{-i}\right),X\right)$,
so $\varphi$ satisfies TI.
\end{proof}

\begin{proof}[\textbf{Proof of Theorem~\ref{thm:characterization extension}:}]
Without loss of generality, for the duration of this proof we assume that $\pi$ is the identity priority. We start with the \textbf{``if''} direction. First, it is clear that $\varphi^{\pi}$ satisfies NW\textsuperscript{*}, RP-$\pi$, and IR.

\textbf{(EF1)} To see that $\varphi^{\pi}$ satisfies EF1, consider
any problem $\left(\succeq,X\right)$ and distinct agents $i,j\in N$
with $i<j$. It suffices to exhibit a set $S\subseteq\varphi_{i}^{\pi}\left(\succeq,X\right)$
such that $\lvert S\rvert \leq1$ and $\varphi_{j}^{\pi}\left(\succeq,X\right)\succeq_{j}^{PD}\varphi_{i}^{\pi}\left(\succeq,X\right)\setminus S$. If $\varphi_{j}^{\pi}\left(\succeq,X\right)=\emptyset$, then the
definition of $\varphi^{\pi}$ implies that $\lvert \varphi_{i}^{\pi}\left(\succeq,X\right) \cap U(\succeq_{j}) \rvert \leq 1$;
hence, $S=\varphi_{i}^{\pi}\left(\succeq,X\right) \cap U(\succeq_j)$ will do.

Assume that
$\ell \coloneqq |\varphi_{j}^{\pi}\left(\succeq,X\right)| \geq 1$. We further assume that $\varphi_{j}^{\pi}\left(\succeq,X\right) \nsucceq^{PD}_j \varphi^\pi_i(\succeq,X)$ (otherwise we may take $S = \emptyset$). Then the definition of $\succeq^{PD}_j$ implies that $j$ finds at least one of $i$'s assigned objects acceptable, i.e., $k \coloneqq |\varphi^\pi_i(\succeq,X) \cap U(\succeq_j)| \geq 1$. Write $\varphi^\pi_i(\succeq,X) \cap U(\succeq_j) \coloneqq  \{x_1 \succ_i \cdots \succ_i x_k\}$ and $\varphi^\pi_j(\succeq,X) = \{y_1 \succ_j \cdots \succ_jy_\ell \}$. Then the definition of $\varphi^\pi$ implies that $k \leq \ell + 1$ (otherwise, $k \geq \ell + 2$ would mean that $j$ was assigned the null object $\omega$ in round~$\ell + 1$ even though object $x_{\ell + 2} \in U(\succeq_j)$ was still available). 

In the draft procedure
\noun{U-Draft}$\left(\succeq,X,\pi\right)$, for each $t=2,\dots,k$,
object $y_{t-1}$ was assigned to agent $j$ before object $x_{t}$
was assigned to agent $i$; hence, $y_{t-1}\succ_{j}x_{t}$ for each
$t=2,\dots,k$. Consequently, $\varphi^{\pi}_{j}\left(\succeq,X\right)\succeq_{j}^{PD}\varphi^{\pi}_{i}\left(\succeq,X\right)\setminus\left\{ x_{1}\right\} $,
which means that EF1 holds.

\textbf{(RM)}
By transitivity of $\succeq_{i}^{PD}$ it suffices to
show that, given a problem $\left(\succeq,X\right)$ and a real object
$x\in\mathbb{O}\setminus X$, one has $\varphi_{i}^{\pi}\left(\succeq,X\cup\left\{ x\right\} \right)\succeq_{i}^{PD}\varphi_{i}^{\pi}\left(\succeq,X\right)$
for each $i\in N$. Suppose $\left(s_{k}\right)_{k=1}^{K}$
is the sequence of selections associated with \noun{U-Draft}$\left(\succeq,X,\pi\right)$,
and $\left(s'_{k}\right)_{k=1}^{K'}$ is the sequence
of selections associated with \noun{U-Draft}$\left(\succeq,X\cup\left\{ x\right\} ,\pi\right)$.
We argue by induction that $X\cup\left\{ x\right\} \setminus\left\{ s'_{1},\dots,s'_{k}\right\} \supseteq X\setminus\left\{ s_{1},\dots,s_{k}\right\} $
for each $k=0,1,\dots,K-1$ (and hence $K' \geq K$).

\textbf{Base case $\left(k=0\right)$. }Clearly, $X\cup\left\{ x\right\} \setminus\emptyset\supseteq X\setminus\emptyset$.

\textbf{Inductive step. }Suppose $X\cup\left\{ x\right\} \setminus\left\{ s'_{1},\dots,s'_{k-1}\right\} \supseteq X\setminus\left\{ s_{1},\dots,s_{k-1}\right\} $
for some $k\in\left\{ 1,\dots,K-1\right\} $. We must show that $X\cup\left\{ x\right\} \setminus\left\{ s'_{1},\dots,s'_{k}\right\} \supseteq X\setminus\left\{ s_{1},\dots,s_{k}\right\} $.
Let $y\in X\setminus\left\{ s_{1},\dots,s_{k}\right\} $. The induction
hypothesis implies that $X\setminus\left\{ s_{1},\dots,s_{k}\right\} \subseteq X\setminus\left\{ s_{1},\dots,s_{k-1}\right\} \subseteq X\cup\left\{ x\right\} \setminus\left\{ s'_{1},\dots,s'_{k-1}\right\} $,
which means that $y\in X\cup\left\{ x\right\} \setminus\left\{ s'_{1},\dots,s'_{k-1}\right\} $.
Let $i\coloneqq f^\pi(k)$ be the agent who is assigned an object in step $k$ of the
draft procedure. Then
\[
s'_{k}=\operatorname{top}_{\succeq_{i}}\left(X\cup\left\{ x\right\} \setminus\left\{ s'_{1},\dots,s'_{k-1}\right\} \right)\succeq_{i}\operatorname{top}_{\succeq_{i}}\left(X\setminus\left\{ s_{1},\dots,s_{k-1}\right\} \right)=s_{k}\succ_{i}y,
\]
which means that $s'_{k}\succ_{i}y$. Hence, $y\in X\cup\left\{ x\right\} \setminus\left\{ s'_{1},\dots,s'_{k}\right\} $,
as we needed to show.

Since $X\cup\left\{ x\right\} \setminus\left\{ s'_{1},\dots,s'_{k}\right\} \supseteq X\setminus\left\{ s_{1},\dots,s_{k}\right\} $
for each $k=0,1,\dots,K-1$, we have $s'_{k}\succeq_{f^{\pi}\left(k\right)}s_{k}$
for each $k=1,\dots,K$. Therefore, RM holds.

\textbf{(TP)} Consider any problem $\left(\succeq,X\right)$. To see
that $\varphi^{\pi}$ satisfies TP, let $\succeq'_{i}$ be a truncation
of $\succeq_{i}$. The definition of $\varphi^{\pi}$ and the fact
that $U\left(\succeq'_{i}\right)\subseteq U\left(\succeq_{i}\right)$
imply that
\[
\varphi_{i}^{\pi}\left(\succeq,X\right)\cap U\left(\succeq'_{i}\right)=\varphi_{i}^{\pi}\left(\left(\succeq'_{i},\succeq_{-i}\right),X\right).
\]
Therefore, $\varphi_{i}^{\pi}\left(\succeq,X\right)\succeq_{i}^{PD}\varphi_{i}^{\pi}\left(\left(\succeq'_{i},\succeq_{-i}\right),X\right)$
and TP holds.

\textbf{(EP)} Consider any problem $\left(\succeq,X\right)$. To see
that $\varphi^{\pi}$ satisfies EP, let $\succeq'_{i}$ be an extension
of $\succeq_{i}$. The definition of $\varphi^{\pi}$ and the fact
that $U\left(\succeq_{i}\right)\subseteq U\left(\succeq'_{i}\right)$
imply that
\[
\varphi_{i}^{\pi}\left(\succeq,X\right)=\varphi_{i}^{\pi}\left(\left(\succeq'_{i},\succeq_{-i}\right),X\right)\cap U\left(\succeq_{i}\right).
\]
Therefore, $\varphi_{i}^{\pi}\left(\succeq,X\right)\succeq_{i}^{PD}\varphi_{i}^{\pi}\left(\left(\succeq'_{i},\succeq_{-i}\right),X\right)$
and EP holds.

\textbf{(TI)} As $\varphi^\pi$ satisfies IR, TP, and EP, Proposition~\ref{Theorem 5-1} implies that it also satisfies TI.

To establish the converse, we will make use of the following lemma, which is the analog of Lemma~\ref{lem:RM lemma} in the present setting.

\begin{lem}
\label{RM Lemma}Suppose $\varphi$ is an allocation rule satisfying
RM. Consider a problem $\left(\succeq,X\right)$ such that $\varphi\left(\succeq,X\right)=\varphi^{\pi}\left(\succeq,X\right)$
for some priority $\pi$. If $x\in\mathbb{O}\setminus X$ is an object
satisfying
\[
\text{for each }i\in N,\quad\left(y\succ_{i}x\quad\text{for each }y\in\varphi_{i}\left(\succeq,X\right)\right),
\]
then $\varphi_{i}\left(\succeq,X\right)\subseteq\varphi_{i}\left(\succeq,X\cup\left\{ x\right\} \right)$
for each $i\in N$.
\end{lem}
\begin{subproof}
Let $\left(s_{k}\right)_{k=1}^{K}$ be the sequence of selections
associated with the draft procedure \noun{U-Draft}$\left(\succeq,X,\pi\right)$. We
show by induction that, for each $k=1,\dots,K$, $s_{k}\neq\omega$
implies that $s_{k}\in\varphi_{f^\pi\left(k\right)}\left(\succeq,X\cup\left\{ x\right\} \right)$.

\textbf{Base case $\left(k=1\right)$.} If $s_{1}\neq\omega$, then
$s_{1}\in\varphi_{f^\pi\left(1\right)}\left(\succeq,X\right)$. Since
$s_{1}=\operatorname{top}_{\succeq_{f^\pi\left(1\right)}}\left(X\right)$, RM implies
that $s_{1}\in\varphi_{f^\pi\left(1\right)}\left(\succeq,X\cup\left\{ x\right\} \right)$.

\textbf{Inductive step.} Suppose that, for some $k\in\left\{ 2,\dots,K\right\} $,
$s_{\ell}\in\varphi_{f^\pi\left(\ell\right)}\left(\succeq,X\cup\left\{ x\right\} \right)$
whenever $s_{\ell}\neq\omega$ and $1\leq\ell<k$. We must show that
$s_{k}\neq\omega$ implies that $s_{k}\in\varphi_{f^\pi\left(k\right)}\left(\succeq,X\cup\left\{ x\right\} \right)$.

Observe that the inductive hypothesis implies that $\varphi_{i}\left(\succeq,X\right)\cap\left\{ s_{1},\dots,s_{k-1}\right\} =\varphi_{i}\left(\succeq,X\cup\left\{ x\right\} \right)\cap\left\{ s_{1},\dots,s_{k-1}\right\} $
for each $i\in N$. If $s_{k}\neq\omega$, then $s_{k}\in\varphi_{f^\pi\left(k\right)}\left(\succeq,X\right)$.
Since $s_{k}=\operatorname{top}_{\succeq_{f^\pi\left(k\right)}}\left(X\setminus\left\{ s_{1},\dots,s_{k-1}\right\} \right)$,
RM implies that $s_{k}\in\varphi_{f^\pi\left(k\right)}\left(\succeq,X\cup\left\{ x\right\} \right)$.

It follows from the principle of induction that, for each $k=1,\dots,K$,
$s_{k}\neq\omega$ implies that $s_{k}\in\varphi_{f^\pi\left(k\right)}\left(\succeq,X\cup\left\{ x\right\} \right)$.
Hence, $\varphi_{i}\left(\succeq,X\right)\subseteq\varphi_{i}\left(\succeq,X\cup\left\{ x\right\} \right)$
for each $i\in N$.
\end{subproof}

To show the \textbf{``only if''} direction, suppose that $\varphi$ satisfies
wRP\textsuperscript{*}-$\pi$, wEF1\textsuperscript{*}, NW\textsuperscript{*}, RM, IR, and TI. We must show that $\varphi=\varphi^{\pi}$. Toward contradiction, assume that $\varphi\neq\varphi^{\pi}$. Let
\[
\mathcal{C}\coloneqq\left\{ \left(\succeq,X\right)\in\mathcal{R}^{N}\times\mathcal{X}\mid\varphi\left(\succeq,X\right)\neq\varphi^{\pi}\left(\succeq,X\right)\right\} 
\]
be the set of conflict problems, which is nonempty by hypothesis.

We first note that every conflict problem has at least two available objects. Indeed, let $(\succeq,X)$ be a problem with $X = \{x\}$. If $x \notin U(\succeq)$, then IR implies that $\varphi(\succeq,X) = (\emptyset)_{i \in N}$, which coincides with $\varphi^\pi(\succeq,X)$. If $x \in U(\succeq)$, let $i^*$ be the highest-priority agent such that $x \in U(\succeq_{i^*})$. Then NW\textsuperscript{*} and IR imply that $x$ is assigned to some agent who finds it acceptable, while wRP\textsuperscript{*}-$\pi$ implies that it must be assigned to $i^*$. Therefore, $\varphi(\succeq,X)$  and $\varphi^\pi(\succeq,X)$ both assign $x$ to agent~$i^*$, and hence $\varphi(\succeq,X) = \varphi^\pi(\succeq,X)$.

For each problem $\left(\succeq,X\right)$,
define
\[
\sigma\left(\succeq,X\right)\coloneqq\sum_{i\in N}\left|U\left(\succeq_{i}\right)\cap X\right|.
\]
Choose a conflict problem $\left(\succeq,X\right)\in\mathcal{C}$
such that
\begin{enumerate}
\item $\left|X\right|$ is minimal among all problems in $\mathcal{C}$:
for all $\left(\succeq',X'\right)\in\mathcal{C}$, $\left|X\right|\leq\left|X'\right|$;
and
\item $\sigma\left(\succeq,X\right)$ is maximal among all problems in $\mathcal{C}$
with $\left|X\right|$ objects available: for all $\left(\succeq',X'\right)\in\mathcal{C}$,
$\left|X\right|=\left|X'\right|$ implies $\sigma\left(\succeq,X\right)\geq\sigma\left(\succeq',X'\right)$.
\end{enumerate}
Let $\left(s_{k}\right)_{k=1}^{K}$ be the sequence of selections
generated by \noun{U-Draft}$\left(\succeq,X,\pi\right)$. Let $S_{0}\coloneqq\emptyset$
and, for each $k=1,\dots,K$, denote $S_{k}\coloneqq\left\{ s_{1},\dots,s_{k}\right\} \setminus\left\{ \omega\right\}$. Since $\left(\succeq,X\right)\in\mathcal{C}$, we have $\varphi\left(\succeq,X\right)\neq\varphi^{\pi}\left(\succeq,X\right)$. Let $L$ be the earliest step of \noun{U-Draft}$\left(\succeq,X,\pi\right)$ such that $s_{L}\neq\omega$ and $s_{L}\notin\varphi_{f^{\pi}\left(L\right)}\left(\succeq,X\right)$. Denote $i\coloneqq f^{\pi}\left(L\right)$.

\begin{claim}
\label{claim:First divergence happens at the last step}$S_{L}=X$.
Hence all objects in $X$ are assigned at $\varphi^{\pi}\left(\succeq,X\right)$
and $\varphi\left(\succeq,X\right)$.
\end{claim}
\begin{proof}
By definition, $S_{L}\subseteq X$. Suppose $S_{L}\subsetneq X$,
and consider the problem $\left(\succeq,S_{L}\right)$. Since $\left|S_{L}\right|<\left|X\right|$,
the choice of $\left(\succeq,X\right)$ implies $\varphi\left(\succeq,S_{L}\right)=\varphi^{\pi}\left(\succeq,S_{L}\right)$.
In particular, $s_{L}\in\varphi_{i}\left(\succeq,S_{L}\right)$. Now
add all objects in $X\setminus S_{L}$, one at a time, according
to the order in which they are selected during \noun{U-Draft}$\left(\succeq,X,\pi\right)$,
with all objects that are never selected added at the very end. A
recursive application of Lemma~\ref{RM Lemma} implies that $s_{L}\in\varphi_{i}\left(\succeq,X\right)$,
a contradiction. Thus $S_{L}=X$, which implies all objects in $X$
are assigned at $\varphi^{\pi}\left(\succeq,X\right)$.

Because all objects are assigned at $\varphi^{\pi}\left(\succeq,X\right)$
and $\varphi^{\pi}$ satisfies IR, we have $X\subseteq U\left(\succeq\right)$
and hence $U\left(\succeq\right)\cap X=X$. Since $\varphi$ satisfies
NW\textsuperscript{*}, it follows that all objects in $X$ are assigned at $\varphi\left(\succeq,X\right)$.
\end{proof}
By Claim \ref{claim:First divergence happens at the last step}, step
$L$ coincides with the last step at which a real object is assigned
during \noun{U-Draft}$\left(\succeq,X,\pi\right)$, i.e., $L=\max\left\{ k\in\left\{ 1,\dots,K\right\} \mid s_{k}\neq\omega\right\} .$
Hence $S_{L-1}=X\setminus\left\{ s_{L}\right\}$, which is nonempty because $|X| \geq 2$. Consider the problem
$\left(\succeq,S_{L-1}\right)$.

\paragraph{Step 1: Relating $\varphi\left(\succeq,S_{L-1}\right)$ and $\varphi\left(\succeq,X\right)$.}

Since $\left|S_{L-1}\right|<\left|X\right|$, the choice of $\left(\succeq,X\right)$
implies that 
\begin{equation}
\varphi\left(\succeq,S_{L-1}\right)=\varphi^{\pi}\left(\succeq,S_{L-1}\right).\label{eq:Correct allocation at L-1}
\end{equation}

Furthermore, Lemma~\ref{RM Lemma} implies that 
\begin{equation}\label{RM inclusion}
\varphi_{j}\left(\succeq,S_{L-1}\right)\subseteq\varphi_{j}\left(\succeq,X\right)\text{ for all }j\in N.
\end{equation}
Claim~\ref{claim:First divergence happens at the last step} implies all objects are assigned at $\varphi\left(\succeq,X\right)$, thus by (\ref{RM inclusion}) there
is an agent $i^{*}\in N$ such that
\begin{equation}\label{eq:s_L assigned to i*}
\varphi_{i^{*}}\left(\succeq,X\right)=\varphi_{i^{*}}\left(\succeq,S_{L-1}\right)\cup\left\{ s_{L}\right\} ,\qquad\varphi_{j}\left(\succeq,X\right)=\varphi_{j}\left(\succeq,S_{L-1}\right)\text{ for all }j\in N\setminus\left\{ i^{*}\right\} .
\end{equation}
Combining (\ref{eq:Correct allocation at L-1}) and (\ref{eq:s_L assigned to i*})
yields
\begin{equation}
\varphi_{i^{*}}\left(\succeq,X\right)=\varphi_{i^{*}}^{\pi}\left(\succeq,S_{L-1}\right)\cup\left\{ s_{L}\right\} ,\qquad\varphi_{j}\left(\succeq,X\right)=\varphi_{j}^{\pi}\left(\succeq,S_{L-1}\right)\text{ for all }j\in N\setminus\left\{ i^{*}\right\} .\label{eq:another i* relation}
\end{equation}
Since $s_{L}\notin\varphi_{i}\left(\succeq,X\right)$ by the definition
of $L$, we have $i^{*}\neq i$. 

\paragraph{Step 2: The case $X\subseteq U\left(\succeq_{i}\right)$.}

Suppose $X\subseteq U\left(\succeq_{i}\right)$, i.e., agent $i$
finds every object in $X$ acceptable. Since $s_{L}$ is assigned
to agent $i^{*}$ at $\varphi\left(\succeq,X\right)$, IR implies
that $s_{L}\in U\left(\succeq_{i^{*}}\right)$. Because $s_{L}$ remains
available until step $L$ of \noun{U-Draft}$\left(\succeq,X,\pi\right)$,
neither $i$ nor $i^{*}$ were ever assigned $\omega$ before step
$L$. Hence, at $\varphi^{\pi}\left(\succeq,S_{L-1}\right)$, agents
$i$ and $i^{*}$ receive bundles with cardinality equal to the number
of their turns before step $L$. Consider two subcases.
\begin{casenv}
\item If $i^{*}\mathrel{\pi}i$, then $i^{*}$ makes one additional selection
before step $L$ of \noun{U-Draft}$\left(\succeq,X,\pi\right)$, so
$\left|\varphi_{i^{*}}^{\pi}\left(\succeq,S_{L-1}\right)\right|=\left|\varphi_{i}^{\pi}\left(\succeq,S_{L-1}\right)\right|+1$.
Therefore, (\ref{eq:another i* relation}) yields
\[
\left|\varphi_{i^{*}}\left(\succeq,X\right)\right|=\left|\varphi_{i^{*}}^{\pi}\left(\succeq,S_{L-1}\right)\right|+1=\left|\varphi_{i}^{\pi}\left(\succeq,S_{L-1}\right)\right|+2=\left|\varphi_{i}\left(\succeq,X\right)\right|+2.
\]
Since $\varphi_{i^{*}}\left(\succeq,X\right)\subseteq X\subseteq U\left(\succeq_{i}\right)$,
this violates wEF1\textsuperscript{*} at $\left(\succeq,X\right)$. 
\item If $i\mathrel{\pi}i^{*}$, then $i$ and $i^{*}$ make the same number
of selections before step $L$ of \noun{U-Draft}$\left(\succeq,X,\pi\right)$,
so $\left|\varphi_{i^{*}}^{\pi}\left(\succeq,S_{L-1}\right)\right|=\left|\varphi_{i}^{\pi}\left(\succeq,S_{L-1}\right)\right|$.
Therefore, (\ref{eq:another i* relation}) yields
\[
\left|\varphi_{i^{*}}\left(\succeq,X\right)\right|=\left|\varphi_{i^{*}}^{\pi}\left(\succeq,S_{L-1}\right)\right|+1=\left|\varphi_{i}^{\pi}\left(\succeq,S_{L-1}\right)\right|+1=\left|\varphi_{i}\left(\succeq,X\right)\right|+1.
\]
Since $\varphi_{i^{*}}\left(\succeq,X\right)\subseteq X\subseteq U\left(\succeq_{i}\right)$,
this violates wRP\textsuperscript{*}-$\pi$ at $\left(\succeq,X\right)$.
\end{casenv}
In either subcase, we obtain a contradiction.

\paragraph{Step 3: The case $X\nsubseteq U\left(\succeq_{i}\right)$.}

Suppose $X\nsubseteq U\left(\succeq_{i}\right)$, i.e., agent $i$
does not find every object in $X$ acceptable. Then $U\left(\succeq_{i}\right)\cap X\subsetneq X$. 

Let $\succeq'_{i}$ denote the complete extension of $\succeq_{i}$,
and let $\succeq'\coloneqq\left(\succeq'_{i},\succeq_{-i}\right)$.
Since $X\subseteq U\left(\succeq'_{i}\right)$ and $U\left(\succeq_{i}\right)\cap X\subsetneq X$,
we have $\sigma\left(\succeq',X\right)>\sigma\left(\succeq,X\right)$.
By the choice of $\left(\succeq,X\right)$, it follows that $\left(\succeq',X\right)\notin\mathcal{C}$
and hence 
\begin{equation}
\varphi\left(\succeq',X\right)=\varphi^{\pi}\left(\succeq',X\right).\label{eq:Complete extension yields the draft assignment}
\end{equation}

\begin{claim}
\label{claim:Draft at the extension coincides with draft}$\varphi^{\pi}\left(\succeq',X\right)=\varphi^{\pi}\left(\succeq,X\right)$.
\end{claim}
\begin{subproof}
Let $\left(s'_{k}\right)_{k=1}^{K'}$ be the sequence of selections
generated by \noun{U-Draft}$\left(\succeq',X,\pi\right)$. We show
by induction that $s'_{q}=s_{q}$ for each $q=1,\dots,L$.

Let $q\in\left\{ 1,\dots,L\right\} $ and suppose that $s'_{\ell}=s_{\ell}$
for each $\ell\in\left\{ 1,\dots,q-1\right\} $. Then, in each of
the two runs \noun{U-Draft}$\left(\succeq,X,\pi\right)$ and \noun{U-Draft}$\left(\succeq',X,\pi\right)$,
the set of remaining objects at step $q$ is $X\setminus S_{q-1}$. 

If $f^{\pi}\left(q\right)\neq i$, then agent $f^{\pi}\left(q\right)$
has the same preferences in $\succeq$ and in $\succeq'$, and hence
$s'_{q}=s_{q}$.

Suppose that $f^{\pi}\left(q\right)=i$. Since $q\leq L$ and $s_{L}$
is not selected before step $L$, we have $s_{L}\in X\setminus S_{q-1}$.
Moreover, $s_{L}\in U\left(\succeq_{i}\right)$, and since $s_{q}=\text{top}_{\succeq_{f^{\pi}\left(q\right)}}\left(X\setminus S_{q-1}\right)$
we must have $s_{q}\in U\left(\succeq_{i}\right)$ also. Since $\succeq'_{i}$
is the complete extension of $\succeq_{i}$, it ranks every object
in $U\left(\succeq'_{i}\right)\setminus U\left(\succeq_{i}\right)$
below every object in $U\left(\succeq_{i}\right)$. Hence
\[
s_{q}=\text{top}_{\succeq_{i}}\left(X\setminus S_{q-1}\right)=\text{top}_{\succeq'_{i}}\left(X\setminus S_{q-1}\right)=s'_{q}.
\]

Thus $\left(s'_{q}\right)_{q=1}^{L}=\left(s_{q}\right)_{q=1}^{L}$.
Since $S_{L}=X$, each real object in $X$ is selected at the same
step by the same agent in the two runs, which implies $\varphi^{\pi}\left(\succeq',X\right)=\varphi^{\pi}\left(\succeq,X\right)$.
\end{subproof}
Combining (\ref{eq:Complete extension yields the draft assignment})
and Claim \ref{claim:Draft at the extension coincides with draft},
we obtain
\begin{equation}
\varphi\left(\succeq',X\right)=\varphi^{\pi}\left(\succeq,X\right).\label{eq:Final equality}
\end{equation}
In particular, 
\[
\varphi_{i}\left(\succeq',X\right)=\varphi_{i}^{\pi}\left(\succeq,X\right)\subseteq U\left(\succeq_{i}\right),
\]
where the inclusion follows from the fact that $\varphi^{\pi}$ satisfies
IR. Since $\succeq_{i}$ is a truncation of $\succeq'_{i}$, TI implies
that 
\[
\varphi_{i}\left(\succeq,X\right)=\varphi_{i}\left(\succeq',X\right).
\]
By (\ref{eq:Final equality}), it follows that $\varphi_i(\succeq, X) = \varphi^\pi_i(\succeq,X)$. But $s_{L}\in\varphi_{i}^{\pi}\left(\succeq,X\right)$ implies that
$s_{L}\in\varphi_{i}\left(\succeq,X\right)$, contradicting the choice
of $L$.
\end{proof}

\section*{Acknowledgements}

We thank Haris Aziz, Jeff Borland, Haluk Ergin, Onur Kesten, Fuhito Kojima, Maciej Kotowski, Simon Loertscher, Robert Macdonald, Alexandru Nichifor, Szilvia P{\'a}pai, Alex Teytelboym, Josh Vanderloo, and Steven Williams for helpful feedback. We are especially grateful to two anonymous reviewers and the Associate Editor, whose detailed comments greatly improved the manuscript. We also thank conferences and seminar audiences at the 2022 SAET Conference, 2023 Deakin Economic Theory Workshop, 3rd Padua Meeting of Economic Design, Australian National University, Cairns Workshop on Market Design, Monash, NES, Notre Dame, SLMath, and University of Queensland. Coreno gratefully acknowledges financial support from an Australian Government RTP Scholarship, a University of Melbourne FBE-GRATS Scholarship, an M. A. Bartlett Scholarship, and the Swiss National Science Foundation through Project 100018-231777. This material is based upon work supported by the National Science Foundation under Grant No.\ DMS-1928930, while Coreno was in residence at the Simons Laufer Mathematical Sciences Institute, Fall 2023.

\bibliographystyle{econ-econometrica}
\phantomsection\addcontentsline{toc}{section}{\refname}\bibliography{Draft_bibliography_new_abbrev}

\end{document}